# Photometric study and absolute parameters estimation of six totally eclipsing contact binaries

Kai Li[1], Qi-Qi Xia[1], Chun-Hwey Kim[2], Xing Gao[3], Shao-Ming Hu[1], Di-Fu Guo[1], Dong-Yang Gao[1], Xu Chen[1], Ya-Ni Guo[1]

## ABSTRACT

High precision CCD observations of six totally eclipsing contact binaries were presented and analyzed. It is found that only one target is an A-type contact binary (V429 Cam), while the others are W-type contact ones. By analyzing the times of light minima, we discovered that two of them exhibit secular period increase while three manifest long-term period decrease. For V1033 Her, a cyclic variation superimposed on the long-term increase was discovered. By comparing the *Gaia* distances with those calculated by the absolute parameters of 173 contact binaries, we found that *Gaia* distance can be applied to estimate absolute parameters for most contact binaries. The absolute parameters of our six targets were estimated by using their *Gaia* distances. The evolutionary status of contact binaries was studied, we found that the A- and W- subtype contact binaries may have different formation channels. The relationship between the spectroscopic and photometric mass ratios for 101 contact binaries was presented. It is discovered that the photometric mass ratios are in good agreement with the spectroscopic ones for almost all the totally eclipsing systems, which is corresponding to the results derived by Pribulla et al. (2003a) and Terrell & Wilson (2005).

[1]Shandong Key Laboratory of Optical Astronomy and Solar-Terrestrial Environment, School of Space Science and Physics, Institute of Space Sciences, Shandong University, Weihai, Shandong, 264209, China (e-mail: kaili@sdu.edu.cn (Li, K.))

[2]Department of Astronomy and Space Science, Chungbuk National University, Cheongju 361-763, Korea

[3]Xinjiang Astronomical Observatory, 150 Science 1-Street, Urumqi 830011, China

## 1. Introduction

W UMa-type contact binaries are short period eclipsing binaries composing two late type components who are overfilling the inner Roche lobes and sharing a common envelope. They exhibit continuous light variability and nearly equal depths of the two eclipsing minima, revealing that their two component stars have similar temperatures. Despite more than five decades of efforts to comprehend their formation, evolution, structure, and ultimate destiny (e.g., Lucy 1968; Bradstreet & Guinan 1994; Eggleton & Kisseleva-Eggleton 2002; Qian 2003; Yakut & Eggleton 2005; Stepien 2006, 2011; Eggleton 2012), no theoretical model can clearly and superiorly interpret their observed properties. Besides, the 0.22 days short period cut-off (Rucinski 1992; Stepien 2006, 2011; Jiang et al. 2012; Li et al. 2019b), the minimum mass ratio (e.g., Rasio 1995; Li & Zhang 2006; Arbutina 2007; Jiang et al. 2010; Caton et al. 2019), O'Connell effect (e.g., O'Connell 1951; Liu & Yang 2003; Lee et al. 2009; Qian et al. 2014; Zhou et al. 2016a), and the thermal relaxation oscillation theory (e.g., Lucy 1976; Flannery 1976; Robertson & Eggleton 1977; Lucy & Wilson 1979; Qian 2001) are still controversial. Due to the research of Binnendijk (1970), W UMa-type contact binaries can be divided into two subtypes: A-subtype and W-subtype. For A-subtype contact systems, the more massive component has the higher effective temperature, while for W-subtype ones, the less massive component has the higher effective temperature. To better understand these issues, the fundamental parameters (such as masses, radii, and luminosities) of a large number of contact binaries should be determined and analyzed.

To derive the fundamental parameters of a contact binary, both photometric light curves and radial velocity observations are required. For one binary, radial velocity measurements require a much larger telescope comparing to the photometric observations. Therefore, radial velocity measurements have been obtained for only a small amount of contact binaries. The *Gaia* mission (Gaia Collaboration et al. 2016, 2018) provides the opportunity (e.g., Kjurkchieva et al. 2019b) to determine the absolute parameters without radial velocity measurements. This mission has derived very precise distances of 1.3 billion objects (Bailer-Jones et al. 2018). Using the *Gaia* distance, researchers have estimated fundamental parameters for many contact binaries (e.g., Kjurkchieva et al. 2019a, 2020; Li et al. 2019a). This estimation can only be applied to contact binaries with accurate light-curve solutions.

By the statistical work on contact binaries with both spectroscopic and photometric observations, Pribulla et al. (2003a) discovered that the photometric mass ratios of those contact binaries whose light curves have flat-bottom minima are nearly equal to their spectroscopic ones, meaning that contact binaries undergoing total eclipses can be determined credible mass ratios without spectroscopic observations. The numerical simulations carried out by Terrell & Wilson (2005) have confirmed this result. Recently, Zhang et al. (2017)

suggested that the degree of symmetry of the light curves and the sharpness of the q-search curve bottom can be used to estimate the reliability of the light-curve solutions. The solution results would be reliable if the light curves are symmetric and the q-search curve bottom is very sharp. In conclusion, accurate and reliable physical parameters can be determined by the investigation of photometric light curves only for totally eclipsing contact binaries with symmetric light curves.

In this paper, we present light-curve solutions for six totally eclipsing contact binaries with symmetric light curves. The information of the six targets are displayed in Table 1. Their absolute parameters were estimated by using *Gaia* distances, and then the evolutionary status of contact binaries were analyzed and summarized.

## 2. Observations and the light-curve solutions of six contact binaries

We carried out observations on the six totally eclipsing contact binaries using the Weihai Observatory 1.0-m telescope of Shandong University (WHOT, Hu et al. 2014), the 60cm Ningbo Bureau of Education and Xinjiang Observatory Telescope (NEXT), and the 60 cm telescope at the Xinglong Station of National Astronomical Observatories (XL60). The PIXIS 2048B camera, the FLI PL23042 camera, and the Andor DU934P camera were equipped on WHOT, NEXT and XL60, respectively. The observational log of the six targets is shown in Table 2. The CCD data for all targets were reduced by using the IRAF[1] following the standard process including bias and flat corrections and aperture photometry. Then, the differential magnitudes between the target and the comparison star were derived.

We used the Wilson-Devinney (W-D) code (Wilson & Devinney 1971; Wilson 1979,

---

[1] IRAF is distributed by the National Optical Astronomy Observatories, which are operated by the Association of Universities for Research in Astronomy under cooperative agreement with the National Science Foundation.

Table 1: Information of the six targets

| Star | RA | Dec | Period (d) | $V_{max}$ | Amplitude | $HJD_0$ | References |
|---|---|---|---|---|---|---|---|
| V429 Cam | 06 13 06.61 | +76 29 48.9 | 0.4411603 | 12.931 | 0.43 | 2458837.36891 | (1), (2), (3) |
| V473 Cam | 07 17 04.92 | +77 10 26.0 | 0.2984374 | 11.399 | 0.61 | 2458532.12468 | (1), (2), (3) |
| V505 Cam | 09 08 49.09 | +82 46 05.9 | 0.3363591 | 13.800 | 0.61 | 2458543.25665 | (2), (3), (4) |
| V514 Cam | 11 28 30.17 | +79 38 56.1 | 0.3627369 | 12.249 | 0.57 | 2458215.39161 | (1), (2), (3) |
| V830 Cep | 20 54 43.71 | +69 59 58.5 | 0.2601338 | 13.400 | 0.45 | 2458388.08311 | (2), (3), (4) |
| V1033 Her | 16 50 39.92 | +27 44 23.0 | 0.2980523 | 11.735 | 0.60 | 2458238.29761 | (1), (2), (3) |

(1) Hoffmann 2009; (2) Shappee et al. 2014; (3) Jayasinghe et al. 2018; (4)Samus et al. 2017

1990, 1994) to model their light curves (the $B$ bandpass light curve of V830 Cep was not used during the light-curve solution because of the large observation errors and dispersion, see Table 2 and Figure 2). The temperatures of the six contact binaries do not affect the photometric results much, but they have great influence on the estimation of the fundamental parameters. Therefore, we should determine reliable temperatures of these contact binaries. Different dereddened color indices and the spectra determined by the Large Sky Area Multi-Object Fiber Spectroscopic Telescope (LAMOST, Luo et al. 2015a) were applied to obtain the temperatures of our six targets, and the results are listed in Table 3. Column (2) displays the value of E(B-V), columns (3) to (6) give the temperatures determined from dereddened color indices of $B - V$, $J - K$, and $g' - i'$ and LAMOST spectra, column (7) gives the mean value of temperature (this value has been rounded to two or three digits). The color indices of $B - V$ and $g' - i'$ were taken from AAVSO Photometric All Sky Survey (APASS) DR9 (Henden et al. 2016), while those of $J - K$ were taken from Two Micron All Sky Survey (2MASS) (Cutri et al. 2003). The temperatures based on dereddened $B - V$ and $J - K$ color indices were derived due to Table 5 of Pecaut & Mamajek (2013), while those according to dereddened $g' - i'$ color index were obtained due to Table 3 of Covey et al. (2007). The interstellar extinctions for different filters were derived by using the S & F method (Schlafly & Finkbeiner 2011) from the IRAS database[2]. The average value, $T_m$, was adopted to be used in the light-curve solutions. For an eclipsing binary, the color index used in our temperature estimation is neither the primary component's nor the secondary one's, it is actually some average value of the two components. So this value was initially set as the temperature of the primary, when we obtained the radius ratio, $k = r_2/r_1$ and the temperature ratio, $T_2/T_1$, the final temperature of each component can be computed by using the equation below (Zwitter et al. 2003; Christopoulou & Papageorgiou 2013),

$$T_1 = (((1+k^2)T_m^4)/(1+k^2(T_2/T_1)^4))^{0.25},$$
$$T_2 = T_1(T_2/T_1).$$ (1)

Since there were no radial velocity curves for our 6 targets, the widely used $q$-search method was applied to determine their mass ratios. The step size is 0.01 when $q < 0.5$, while that is 0.05 when $q > 0.5$. During this process, the fitted parameters were the orbital inclination $i$, secondary temperature, $T_2$, the monochromatic luminosity of primary, $L_1$, and the potential ($\Omega = \Omega_1 = \Omega_2$). The limb-darkening coefficients were automatically interpolated from the table of Van Hamme (1993). The $q$-search curves for the six targets are shown in Figure 1. From this figure, we can see very clear minima for all our targets, and all the bottoms are very deep and sharp, revealing that the mass ratio can be well determined for each target. Then, an unambiguous value of mass ratio can be obtained when making it an adjustable parameter, other parameters can be derived accordingly and are shown in Table 4. The errors displayed in Table 4 are formal errors and have no physical meanings, they are underestimated. When we determined the best-fitting light-curve solutions, the temperatures of the two component stars were calculated by using Equation (1) and are listed in Table 5. The best-fitting theoretical light curves labeled with continuous lines for the six targets are plotted in Figure 2. Due to our analysis, five targets are W-subtype contact systems, while only one (V429 Cam) is an A-subtype contact system.

---

[2]https://irsa.ipac.caltech.edu/applications/DUST/

The six systems have been observed by many worldwide photometric surveys, such as SuperWASP (Butters et al. 2010), the All-Sky Automated Survey for SuperNovae (ASAS-SN, Shappee et al. 2014; Jayasinghe et al. 2018), and the Transiting Exoplanet Survey Satellite (TESS) (Ricker et al. 2015). We found that the SuperWASP and TESS data have very high precision. All the six targets have been observed by TESS, three sectors have measured V429 Cam, V473 Cam, V505 Cam, they are Sectors 19, 20, 26, four sectors have measured V514 Cam, they are Sectors 14, 20, 21, 26, seven sectors have measured V830 Cep, they are Sectors 16, 17, 18, 19, 22, 24, 25, while only one sector has measured V1033 Her, it is Sector 25. By visual inspection, we found that some of the data cannot be used for the following light curve solution. They are Sectors 20 and 26 of V429 Cam, Sectors 19, 20, 26 of V505 Cam, Sector 22 of V830 Cep, and Sector 25 of V1033 Her. Only one target, V1033 Her, has been observed by SuperWASP, and the data can be used for the following analysis. If the shapes of the light curves are alike, these light curves were analyzed together, while the light curves exhibiting different shapes were analyzed separately. Because all the TESS data of our targets are in a 30-minute cadence, the phase smearing effect was considered during the light curve solutions (Zola et al. 2017). Therefore, we cannot analyze the TESS data together with our observed data. In addition, the mass ratio which was adopted the value determined by our observations was set as a fixed parameter during the solutions. The results are listed in Table 6, and the corresponding fitted light curves are plotted in Figure 3. We found that the photometric results derived by the TESS and SuperWASP data are very similar to those determined by our observation data.

## 3. Estimation on the absolute parameters of the six contact binaries

### 3.1. Statistics on contact binaries whose absolute parameters have been directly derived

Before using *Gaia* distances to estimate the absolute parameters of our six targets, we carried out a statistics on contact binaries who have been analyzed by both radial velocity

Table 2: Observation log of the six targets

| Star | UT Date yyyymmdd | Exposures s | Mean error mag | Comparison 2MASS | Check 2MASS | Type | Telescope |
|---|---|---|---|---|---|---|---|
| V429 Cam | 2019 Dec 16 | V60 R40 I50 | V0.006 R0.007 I0.006 | 06141310+7632578 | 06113077+7628316 | light curve | NEXT |
| | 2019 Dec 17 | V60 R40 I50 | V0.008 R0.010 I0.011 | 06141310+7632578 | 06113077+7628316 | light curve | NEXT |
| | 2019 Dec 19 | V60 R40 I50 | V0.013 R0.012 I0.013 | 06141310+7632578 | 06113077+7628316 | light curve | NEXT |
| | 2019 Dec 20 | V60 R40 I50 | V0.008 R0.006 I0.006 | 06141310+7632578 | 06113077+7628316 | light curve | NEXT |
| | 2019 Dec 19 | V60 R40 I30 | V0.008 R0.006 I0.007 | 06141310+7632578 | 06113077+7628316 | light curve | WHOT |
| V473 Cam | 2019 Feb 17 | B25 V10 R4 I6 | B0.006 V0.005 R0.004 I0.004 | 07171122+7712384 | 07180929+7714230 | light curve | WHOT |
| V505 Cam | 2019 Feb 17 | V100 R60 I40 | V0.008 R0.006 I0.006 | 09103994+8248219 | 09102259+8246208 | light curve | WHOT |
| V514 Cam | 2015 Apr 25 | B80 V60 R35 I25 | B0.006 V0.005 R0.004 I0.004 | 11294968+7941514 | 11270918+7939280 | minimum | WHOT |
| | 2018 Apr 29 | B60 V40 R20 I30 | B0.010 V0.008 R0.006 I0.009 | 11294968+7941514 | 11270918+7939280 | light curve | NEXT |
| V830 Cep | 2018 Sep 06 | B70 V40 R20 I10 | B0.056 V0.008 R0.007 I0.007 | 20551135+7003013 | 20551123+6957568 | light curve | WHOT |
| V1033 Her | 2015 Apr 25 | R40 | R0.003 | 16504691+2740452 | 16505180+2746463 | minimum | WHOT |
| | 2018 Mar 30 | R30 | R0.008 | 16504691+2740452 | 16505180+2746463 | minimum | WHOT |
| | 2018 Apr 28 | B60 V40 R20 I30 | B0.010 V0.009 R0.008 I0.010 | 16504691+2740452 | 16505180+2746463 | light curve | NEXT |
| | 2018 Apr 29 | B60 V40 R20 I30 | B0.010 V0.008 R0.006 I0.009 | 16504691+2740452 | 16505180+2746463 | light curve | NEXT |
| | 2018 Apr 30 | B60 V40 R20 I30 | B0.004 V0.005 R0.004 I0.006 | 16504691+2740452 | 16505180+2746463 | light curve | NEXT |
| | 2018 May 03 | B60 V40 R20 I30 | B0.015 V0.013 R0.011 I0.013 | 16504691+2740452 | 16505180+2746463 | light curve | NEXT |
| | 2019 Feb 13 | R20 | R0.004 | 16504691+2740452 | 16505180+2746463 | minimum | NEXT |
| | 2020 Feb 19 | R20 | R0.006 | 16504691+2740452 | 16505180+2746463 | minimum | XL60 |

Table 3: The temperatures of the six targets

| Star | E(B-V) | $T_{BV}$ (K) | $T_{JK}$ (K) | $T_{g'i'}$ (K) | $T_{LAMOST}$ (K) | $T_m$ (K) |
|---|---|---|---|---|---|---|
| V429 Cam | 0.114 | 6340 | 6270 | 6400 | − | 6350 |
| V473 Cam | 0.037 | 5430 | 5270 | 5550 | − | 5400 |
| V505 Cam | 0.023 | 5890 | 5520 | 5770 | − | 5750 |
| V514 Cam | 0.051 | 5710 | 5610 | 5770 | − | 5700 |
| V830 Cep | 0.360 | 6520 | 5380 | 6250 | − | 6050 |
| V1033 Her | 0.060 | 5460 | 5450 | 5580 | 5426 | 5500 |

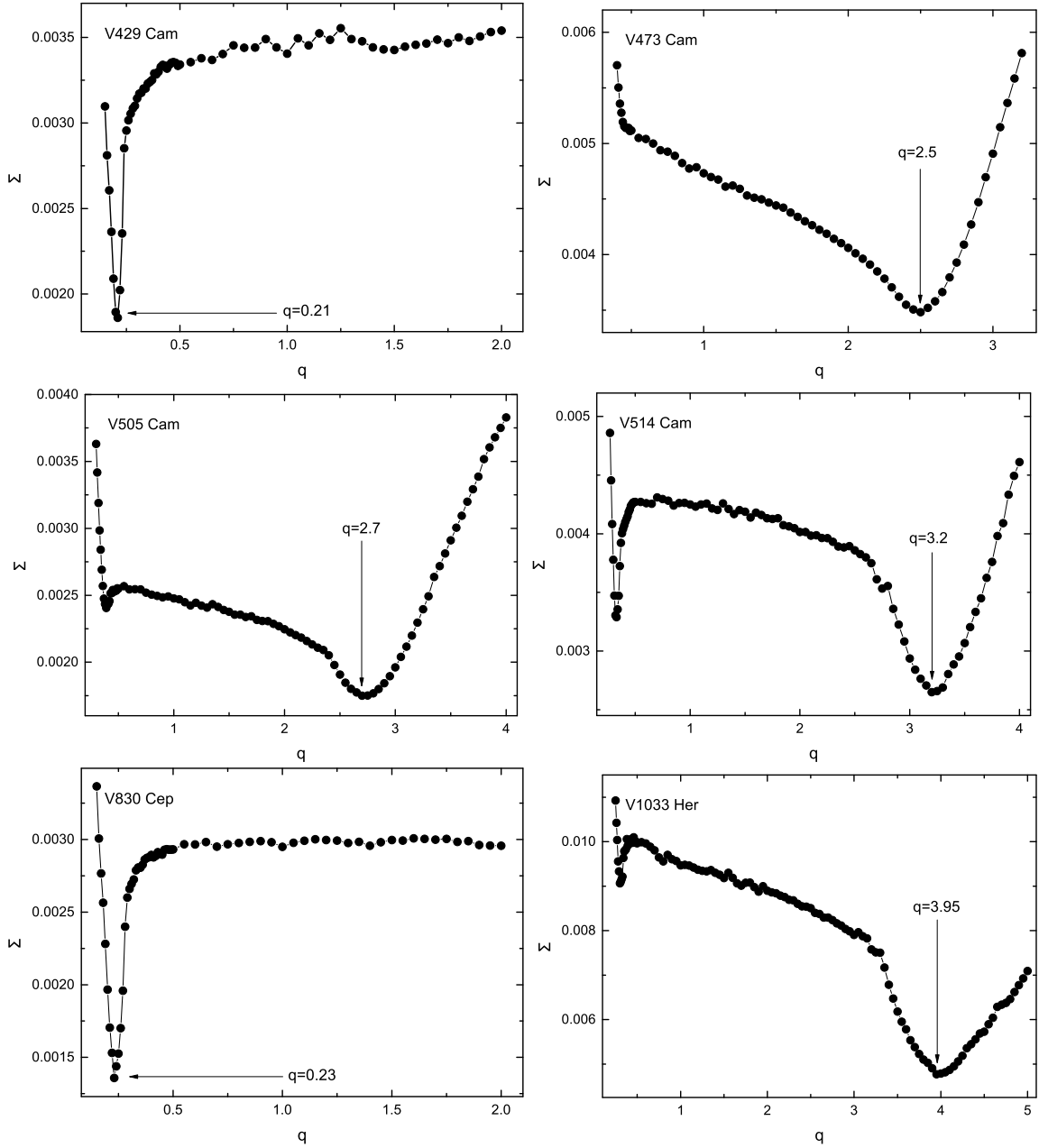

Fig. 1.— The relationship between $\Sigma W_i(O-C)_i^2$ and mass ratio $q$ of the six targets.

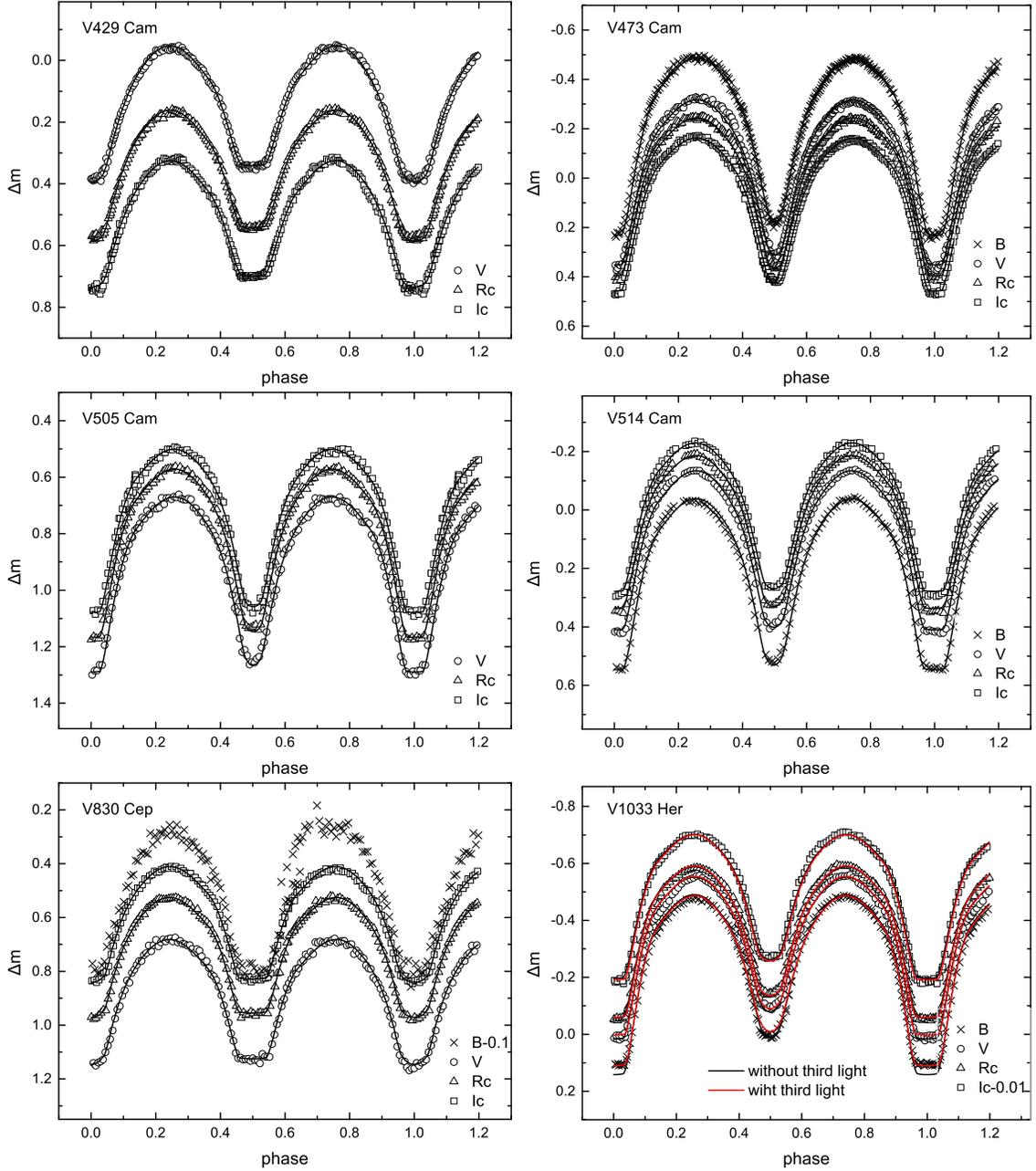

Fig. 2.— The best-fitting theoretical light curves for the six targets.

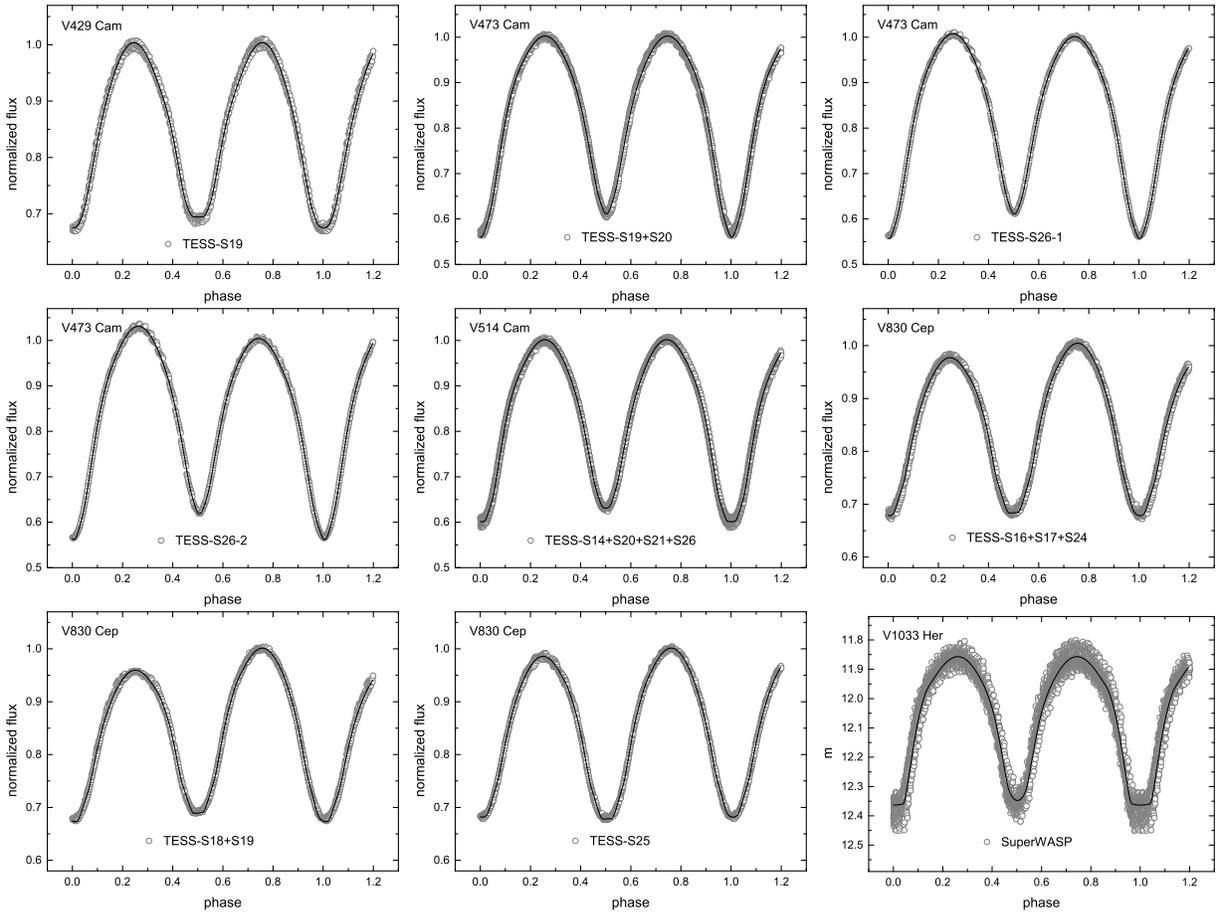

Fig. 3.— The best-fitting theoretical light curves of the TESS and SuperWASP data.

Table 4: Photometric elements of the six contact binaries

| Parameters | V429 Cam | V473 Cam | V505 Cam | V514 Cam | V830 Cep | V1033 Her |
|---|---|---|---|---|---|---|
| $q(M_2/M_1)$ | 0.208±0.001 | 2.492±0.023 | 2.705±0.020 | 3.279±0.022 | 0.228±0.001 | 3.957±0.0093 |
| $T_2/T_1$ | 0.988±0.001 | 0.938±0.001 | 0.964±0.002 | 0.969±0.001 | 1.021±0.001 | 0.926±0.001 |
| $i(deg)$ | 81.5±0.2 | 84.4±0.1 | 85.5±0.2 | 84.3±0.3 | 88.6±0.1 | 84.9±0.2 |
| $\Omega_1 = \Omega_2$ | 2.202±0.004 | 5.849±0.018 | 6.146±0.026 | 6.904±0.028 | 2.262±0.003 | 7.781±0.022 |
| $l_{2B}/l_{1B}$ | – | 1.762±0.006 | – | 2.366±0.012 | – | 2.044±0.009 |
| $l_{2V}/l_{1V}$ | 0.240±0.001 | 1.866±0.006 | 2.053±0.011 | 2.494±0.010 | 0.295±0.001 | 2.301±0.009 |
| $l_{2R_c}/l_{1R_c}$ | 0.242±0.001 | 1.932±0.006 | 2.116±0.010 | 2.560±0.010 | 0.291±0.001 | 2.473±0.010 |
| $l_{2I_c}/l_{1I_c}$ | 0.243±0.001 | 1.979±0.006 | 2.161±0.010 | 2.608±0.009 | 0.287±0.001 | 2.599±0.009 |
| $l_{3B}/l_B$ | – | – | – | – | – | 5.6±0.2% |
| $l_{3V}/l_V$ | – | – | – | – | – | 2.6±0.1% |
| $l_{3R_c}/l_{R_c}$ | – | – | – | – | – | 0.8±0.1% |
| $l_{3I_c}/l_{I_c}$ | – | – | – | – | – | 0 |
| $r_{1(pole)}$ | 0.496 ± 0.001 | 0.290 ± 0.001 | 0.283 ± 0.001 | 0.267 ± 0.001 | 0.484 ± 0.001 | 0.252 ± 0.001 |
| $r_{1(side)}$ | 0.543 ± 0.001 | 0.303 ± 0.001 | 0.295 ± 0.001 | 0.279 ± 0.001 | 0.526 ± 0.001 | 0.263 ± 0.001 |
| $r_{1(back)}$ | 0.569 ± 0.001 | 0.340 ± 0.001 | 0.332 ± 0.001 | 0.316 ± 0.001 | 0.552 ± 0.001 | 0.300 ± 0.001 |
| $r_{2(pole)}$ | 0.249 ± 0.003 | 0.439 ± 0.002 | 0.443 ± 0.002 | 0.461 ± 0.002 | 0.255 ± 0.002 | 0.481 ± 0.001 |
| $r_{2(side)}$ | 0.261 ± 0.003 | 0.470 ± 0.002 | 0.475 ± 0.003 | 0.496 ± 0.003 | 0.269 ± 0.002 | 0.524 ± 0.001 |
| $r_{2(back)}$ | 0.307 ± 0.007 | 0.499 ± 0.003 | 0.503 ± 0.004 | 0.524 ± 0.004 | 0.313 ± 0.004 | 0.551 ± 0.001 |
| $r_1^a$ | 0.537±0.001 | 0.313±0.001 | 0.305±0.001 | 0.289±0.001 | 0.524±0.001 | 0.274±0.001 |
| $r_2^a$ | 0.273±0.004 | 0.471±0.002 | 0.477±0.003 | 0.493±0.003 | 0.274±0.003 | 0.509±0.003 |
| $f$ | 37.9±2.9% | 14.1±2.9% | 12.4±4.2% | 12.6±4.5% | 26.2±2.1% | 12.0±3.5% |

$^a$ Equal-volume radius.

Table 5: The final temperatures of the two component stars for each target

| Star | V429 Cam | V473 Cam | V505 Cam | V514 Cam | V830 Cep | V1033 Her |
|---|---|---|---|---|---|---|
| $T_1$(K) | 6366(23) | 5637(31) | 5897(46) | 5835(47) | 6021(20) | 5828(46) |
| $T_2$(K) | 6287(29) | 5286(34) | 5686(53) | 5652(54) | 6147(28) | 5399(47) |

curves and photometric light curves. Thanks to the seminal works by Rucinski (e.g., Rucinski et al. 2000, 2001, 2002), more than one hundred contact binaries have been observed spectroscopically and photometrically. Through the literature research, we collected contact binaries whose absolute parameters have been well-determined and show them in Table 7. In Table 7, column (1) is the name of the contact binary, column (2) gives the subtype due to the Binnendijk (1970)'s classification, columns (3)-(10) show the period, V magnitude at maximum brightness, spectroscopic and photometric mass ratios, orbital inclination, the fillout factor, and temperatures of the two component stars, columns (11)-(17) display the semi-major axis, masses, radii, and luminosities of the two component stars, columns (18)-(23) list the extinction, bolometric correction, $Gaia$ distance (which is determined by Bailer-Jones et al. 2018), distance from the literature, distance calculated by us, and the reference. In this table, the extinction was determined by using S & F method from the IRAS database, the bolometric correction was interpolated from the the online table [3] provided by Pecaut & Mamajek (2013) due to the temperature of the more massive component. The distance calculated by us is using the equation $d = 10^{\frac{m_{Vmax} - M_V + 5 - A_V}{5}}$, where $A_V$ is the interstellar extinction, $m_{Vmax}$ is the V magnitude at maximum brightness, $M_V$ is the V band absolute magnitude calculated by $M_V = M_{bol} - BC_V$, and $M_{bol} = -2.5 \log(\frac{L_1 + L_2}{L_\odot}) + 4.75$. In total, 173 contact binaries have been collected.

Based on our statistics, the relationship between $Gaia$ distances and those calculated by using the absolute parameters for the 173 contact binaries is shown in the left panel of Figure 4. The right panel is an enlarged figure of the left panel ranging from 0 to 600 pc. The black dots represent our calculated distances, while the red ones show the distances taken from the reference. The solid line is the one-to-one line, while the two dashed lines denote the 25% discrepancy boundary between the two distances. If the distance of one binary has been calculated by the literature, this value will be shown in Figure 4, or the distance calculated by us will be used. From Figure 4, we found that more than eighty percent of the contact binaries are within the 25% boundary. Therefore, we can use $Gaia$ distances to estimate the absolute parameters for most contact binaries. For the 32 contact binaries who have very large discrepancy between the two distances, there are many possible reasons, such as too distant, very large proportion of third light, inaccurate interstellar extinction, inaccurate V band magnitude at maximum brightness.

---

[3]http://www.pas.rochester.edu/~emamajek/EEM_dwarf_UBVIJHK_colors_Teff.txt

Table 6: Photometric elements of five contact binaries using photometric survey data

| Parameters | V429 Cam | V473 Cam | | | V514 Cam | V830 Cep | | | V1033 Her |
|---|---|---|---|---|---|---|---|---|---|
| | S19 | S19+20 | S26-1 | S26-2 | S14+20+21+26 | S16+17+24 | S18+19 | S25 | SuperWASP |
| $q(M_2/M_1)$ | 0.208 | 2.492 | 2.492 | 2.492 | 3.279 | 0.228 | 0.228 | 0.228 | 3.957 |
| $T_2/T_1$ | 0.999±0.001 | 0.944±0.001 | 0.958±0.001 | 0.964±0.001 | 0.943±0.001 | 1.028±0.001 | 1.017±0.001 | 1.036±0.001 | 0.959±0.001 |
| $i(deg)$ | 81.7±0.2 | 82.3±0.1 | 82.1±0.1 | 82.2±0.1 | 85.6±0.3 | 89.6±0.1 | 87.9±0.1 | 89.6±0.1 | 85.9±0.2 |
| $\Omega_1 = \Omega_2$ | 2.172±0.001 | 5.871±0.002 | 5.854±0.002 | 5.834±0.003 | 6.848±0.001 | 2.269±0.001 | 2.274±0.001 | 2.268±0.001 | 7.784±0.002 |
| $l_2/l_1$ | 0.262±0.001 | 1.820±0.002 | 1.929±0.004 | 1.762±0.006 | 2.316±0.002 | 0.293±0.001 | 0.281±0.001 | 0.300±0.001 | 2.784±0.007 |
| $r_{1(pole)}$ | 0.503±0.001 | 0.288±0.001 | 0.289±0.001 | 0.290±0.001 | 0.271 ± 0.001 | 0.485±0.001 | 0.483±0.001 | 0.484±0.001 | 0.253±0.001 |
| $r_{1(side)}$ | 0.554±0.001 | 0.301±0.001 | 0.302±0.001 | 0.305±0.001 | 0.284 ± 0.001 | 0.527±0.001 | 0.526±0.001 | 0.528±0.001 | 0.264±0.001 |
| $r_{1(back)}$ | 0.583±0.001 | 0.337±0.001 | 0.339±0.001 | 0.342±0.001 | 0.324 ± 0.001 | 0.553±0.001 | 0.551±0.001 | 0.553±0.001 | 0.300±0.001 |
| $r_{2(pole)}$ | 0.257±0.001 | 0.437±0.001 | 0.438±0.001 | 0.440±0.001 | 0.463 ± 0.001 | 0.249±0.001 | 0.248±0.001 | 0.250±0.001 | 0.473±0.001 |
| $r_{2(side)}$ | 0.271±0.001 | 0.468±0.001 | 0.469±0.001 | 0.471±0.001 | 0.500 ± 0.001 | 0.261±0.001 | 0.259±0.001 | 0.261±0.001 | 0.512±0.001 |
| $r_{2(back)}$ | 0.331±0.001 | 0.496±0.001 | 0.498±0.001 | 0.501±0.001 | 0.528 ± 0.001 | 0.301±0.001 | 0.298±0.001 | 0.301±0.001 | 0.537±0.001 |
| $r_1{}^a$ | 0.548±0.001 | 0.310±0.001 | 0.312±0.001 | 0.314±0.001 | 0.295±0.001 | 0.522±0.001 | 0.521±0.001 | 0.523±0.001 | 0.274±0.001 |
| $r_2{}^a$ | 0.285±0.001 | 0.468±0.001 | 0.470±0.001 | 0.472±0.001 | 0.498±0.001 | 0.272±0.001 | 0.270±0.001 | 0.272±0.001 | 0.509±0.001 |
| Spot | – | – | Star 2 | Star 2 | – | Star 1 | Star 1 | Star 1 | – |
| $\theta$(radian) | – | – | 0.739 | 0.520 | – | 0.590 | 0.778 | 0.703 | – |
| $\phi$(radian) | – | – | 3.483 | 3.879 | – | 4.908 | 4.958 | 4.817 | – |
| $r$(radian) | – | – | 0.290 | 0.446 | – | 0.401 | 0.405 | 0.251 | – |
| $T_f(T_d/T_0)$ | – | – | 0.840 | 0.879 | – | 0.823 | 0.812 | 0.789 | – |
| $f$ | 60.1±0.9% | 10.4±0.3% | 13.2±0.4% | 16.5±5.5% | 21.6±0.2% | 21.9±0.3% | 18.4±0.4% | 22.4±0.3% | 11.6±0.5% |

$^a$ Equal-volume radius.

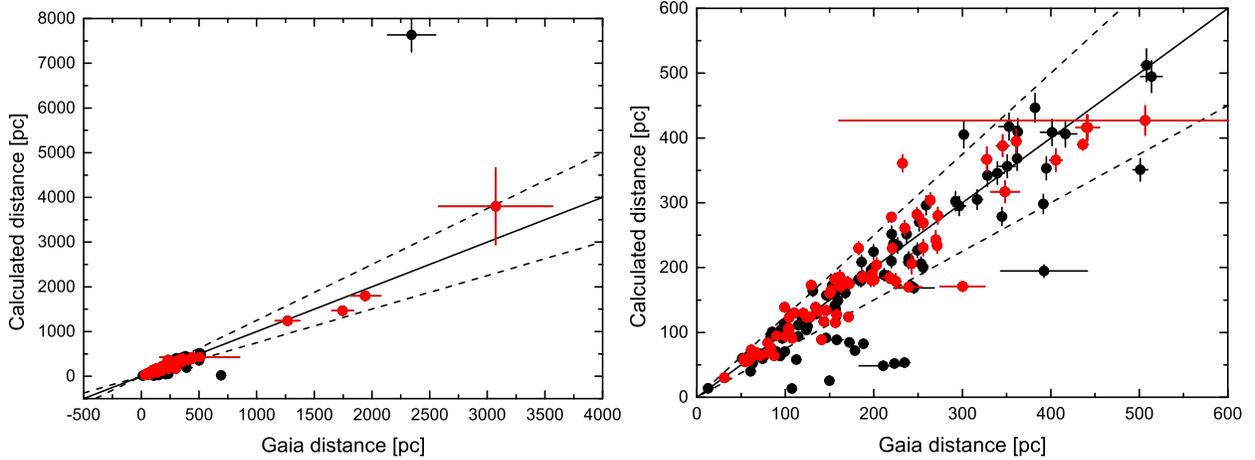

Fig. 4.— The left figure shows the relationship between *Gaia* distance and the distance calculated by using the absolute parameters for the 173 contact binaries, while the right one plots an enlargement of the left panel ranging from 0 to 600 pc. The black dots represent our calculated distances, while the red ones show the distances taken from the reference. The solid line is the one-to-one line, while the two dashed lines denote the 25% discrepancy boundary between the two distances. The errors of *Gaia* distances are obtained from Bailer-Jones et al. (2018), while those of the calculated distances are computed by assuming that the uncertainty of the total luminosity of one binary is one tenth of itself.

Table 7: The 173 contact binaries with spectroscopic and photometric observations

| Name | T | P | Vmax | $q_{sp}$ | $q_{ph}$ | $i$ | $f$ | $T_1$ | $T_2$ | $\alpha$ | $M_1$ | $M_2$ | $R_1$ | $R_2$ | $L_1$ | $L_2$ | $A_V$ | $BC_V$ | $d_G$ | $d_r$ | $d_c$ | Ref. |
|---|---|---|---|---|---|---|---|---|---|---|---|---|---|---|---|---|---|---|---|---|---|---|
| | | days | mag | | mag | ° | % | K | K | $R_\odot$ | $M_\odot$ | $M_\odot$ | $R_\odot$ | $R_\odot$ | $L_\odot$ | $L_\odot$ | mag | mag | pc | pc | pc | |
| AB And | W | 0.33189 | 9.50 | 0.560 | 0.524 | 83 | 25 | 5046 | 5450 | 2.37 | 1.03 | 0.58 | 1.15 | 0.82 | 0.77 | 0.53 | 0.306 | −0.288 | 85 | | 101 | (1), (2), (3) |
| BX And | A | 0.61011 | 8.90 | 0.445 | 0.623 | 76 | 5 | 6650 | 4758 | 4.42 | 2.15 | 0.98 | 2.01 | 1.40 | 7.08 | 0.90 | 0.147 | −0.039 | 182 | | 182 | (4), (5) |
| CN And | A | 0.46279 | 9.70 | 0.390 | 0.450 | 68 | 0 | 6350 | 5732 | 3.14 | 1.40 | 0.55 | 1.45 | 0.94 | 3.41 | 0.93 | 0.170 | −0.049 | 200 | 181 | 193 | (6), (7), (8) |
| GZ And | W | 0.30502 | 10.83 | 0.514 | 0.511 | 86 | 3 | 5021 | 5260 | 2.42 | 1.25 | 0.65 | 1.06 | 0.78 | 0.63 | 0.41 | 0.233 | −0.300 | 168 | | 173 | (9), (10), (11) |
| QX And | A | 0.41182 | 11.25 | 0.306 | 0.233 | 55 | 35 | 6440 | 6420 | 2.89 | 1.47 | 0.45 | 1.46 | 0.88 | 3.19 | 1.15 | 0.150 | −0.042 | 442 | 416 | 395 | (12), (13), (14) |
| V376 And | A | 0.79867 | 7.68 | 0.305 | | 63 | 7 | 9000 | 7080 | 5.32 | 2.44 | 0.74 | 2.60 | 1.51 | 40.0 | 5.00 | 0.760 | −0.122 | 183 | 230 | 193 | (15), (16) |
| V546 And | A | 0.38302 | 11.20 | 0.254 | | 86 | 13 | 6240 | 6517 | 2.46 | 1.08 | 0.28 | 1.23 | 0.66 | 2.05 | 0.71 | 0.178 | −0.060 | 243 | 206 | 306 | (17) |
| OO Aql | A | 0.50679 | 9.20 | 0.846 | 0.843 | 88 | 37 | 6100 | 5926 | 3.35 | 1.90 | 0.90 | 1.41 | 1.31 | 2.45 | 1.89 | 0.064 | −0.078 | 119 | | 122 | (18), (19) |
| V417 Aql | A | 0.37031 | 10.52 | 0.362 | 0.368 | 82 | 24 | 6252 | 6381 | 2.70 | 1.41 | 0.51 | 1.30 | 0.83 | 2.33 | 1.02 | 0.757 | −0.059 | 212 | | 189 | (9), (20), (21) |
| EL Aqr | A | 0.48141 | 10.35 | 0.203 | | 70 | 48 | 6881 | 6137 | 3.25 | 1.66 | 0.34 | 1.76 | 0.90 | 6.25 | 1.03 | 0.076 | −0.015 | 340 | | 346 | (15), (21) |
| HV Aqr | A | 0.37446 | 9.71 | 0.145 | 0.146 | 79 | 56 | 6460 | 6667 | 2.53 | 1.36 | 0.20 | 1.45 | 0.65 | 3.27 | 0.74 | 0.175 | −0.041 | 196 | | 185 | (7), (22), (23) |

(1) Hrivnak 1988, (2) Pych et al. 2004, (3) Li et al. 2014a, (4) Samec et al. 1989a, (5) Siwak et al. 2010, (6) Rafert et al. 1985, (7) Rucinski et al. 2000, (8) Yildirim et al. 2019, (9) Lu & Rucinski 1999, (10) Walker 1973, (11) Yang & Liu 2003a, (12) Qian et al. 2007a, (13) Pribulla et al. 2009a, (14) Djurašević et al. 2011, (15) Rucinski et al. 2001, (16) Çiçek 2011, (17) Gürol et al. 2015b, (18) Hrivnak 1989, (19) Pribulla et al. (2007), (20) Samec et al. 1997, (21) Deb & Singh 2011, (22) Molik & Wolf 2000, (23) Li & Qian 2013, (24) Leung & Schneider 1978, (25) Özkardeş & Erdem 2012, (26) Szalai et al. 2007, (27) Gürol et al. 2015c, (28) Lu 1991, (29) Lin et al. 1993, (30) Kim et al. 2003b, (31) Rucinski & Lu 1999, (32) Vanko et al. 2001, (33) Gazeas et al. 2005, (34) Zola et al. 2004, (35) Rucinski et al. 2003, (36) Yang et al. 2005b, (37) Luo et al. 2017,(38) Hill et al. 1989, (39) Lu et al. 2001, (40) Schieven et al. 1983, (41) Hrivnak 1993, (42) Nelson 2010, (43) Krzesiński et al. 1991, (44) Yang et al. 2012, (45) cSenavcı et al. 2008, (46) Pribulla et al. 2011, (47) Djurašević et al. 2013, (48) Samec et al. 1999, (49) Özdemir et al. 2004, (50) Pribulla et al. 2009b, (51) Christopoulou & Papageorgiou 2013, (52) Rucinski et al. 2005, (53) Gazeas et al. 2006, (54) Gürol et al. 2015a (55) Milone et al. 1991, (56) Samec et al. 1989b, (57) Christopoulou et al. 2012, (58) Hoffmann 1978, (59) Christopoulou & Papageorgiou 2011, (60) Niarchos 1978, (61) Rainger et al. 1990, (62) Liu et al. 2011, (63) McLean & Hilditch 1983, (64) Barone et al. 1993, (65) Baran et al. 2004, (66) Pribulla et al. 2002, (67) Metcalfe 1999, (68) Zola et al. 2001, (69) Liu et al. 2019, (70) Samec et al. 2004, (71) Zhang & Zhang 2004, (72) Noori & Abedi 2017, (73) Twigg 1979, (74) Yang et al. 2005c, (75) Rucinski et al. 1992, (76) Drechsel et al. 1982, (77) Lorenz et al. 1999, (78) Leung 1976, (79) Maceroni et al. 1984, 80) Kaluzny 1984, (81) Mitnyan et al. 2018, (82) Oh et al. 2010, (83) Rucinski et al. 2002, (84) Djurašević et al. 2006, (85) Duerbeck & Rucinski 2007, (86) Russo et al. 1982, (87) Gürol et al. 2016, (88) Hilditch 1981, (89) Pribulla et al. 2006, (90) Zhang et al. 2009, (91) Maceroni et al. 1982, (92) Köse et al. 2011, (93) Zasche & Uhlář 2010, (94) Milone et al. 1987, (95) He & Qian 2008, (96) Nelson & Alton 2019, (97) Goecking & Duerbeck 1993, (98) Alton & Nelson 2018, (99) Essam 2010, (100) Qian et al. 2008, (101) Nelson et al. 2014, (102) Lee et al. 2015, (103) Degirmenci et al. 1999, (104) Harries et al. 1997, (105) Yaşarsoy & Yakut 2013, (106) Wolf et al. 2000, (107) Tian et al. 2018, (108) Rucinski et al. 2008, (109) Ulaş et al. 2012, (110) Ostadnezhad et al. 2014, (111) Kreiner et al. 2003, (112) Sezer et al. 1985, (113) Hilditch et al. 1992, (114) Sarotsakulchai et al. 2019, (115) Kaluzny & Rucinski 1986, (116) Park et al. 2013, (117) Pribulla et al. 2001, (118) Plewa et al. 1991, (119) Yang 2012, (120) Liu et al. 2012, (121) Çalışkan et al. 2014, (122) Papageorgiou & Christopoulou 2015, (123) Liu et al. 1996, (124) Yamasaki et al. 1988, (125) Gu 1999, (126) Mauder 1972, (127) Qian et al. 2007c, (128) Yang & Liu 1999, (129) Nesci et al. 1985, (130) Wang 2017, (131) Yang et al. 2009, (132) Lucy & Wilson 1979, (133) Nelson et al. 1995, (134) Erkan & Ulaş 2016, (135) Erdem &"Özkardeş 2006, (136) Nomen-Torres & Garcia-Melendo 1996, (137) Erdem &"Özkardeş 2009, (138) Özdemir et al. 2002, (139) Selam et al. 2018, (140) Papageorgiou et al. 2015, (141) King & Hilditch 1984, (142) Yang et al. 2004, (143) Qian & Yang 2005, (144) Wadhwa & Zealey 2005, (145) Özkardeş et al. 2009, (146) Alton & Terrell 2006, (147) Kjurkchieva et al. 2018, (148) Binnendijk 1984, (149) Hiller et al. 2004, (150) Gorda 2016, (151) Zhang et al. 1992, (152) Qian et al. 2007b, (153) Vinko et al. 1996, (154) Lee & Park 2018, (155) Barden 1987, (156) Yakut et al. 2003, (157) Niarchos et al. 1994b, (158) Luo et al. 2015b, (159) Liao et al. 2019, (160) Niarchos et al. 1994a, (161) Yang & Liu 2004, (162) Dumitrescu 2003, (163) Kaluzny 1986, (164) Hilditch et al. 1984, (165) Markworth & Michaels 1982, (166) Zola et al. 2005, (167) Lorenzo et al. 2016, (168) Yang et al. 2019, (169) Zhou et al. 2015, (170) Hilditch et al. 1989, (171) Qian et al. 2005, (172) Lapasset & Gomez 1990, (173) Xiang et al. 2015, (174) Degirmenci 2006, (175) Akalin & Derman 1997, (176) Pazhouhesh & Edalati 2002, (177) Rodríguez et al. 1998, (178) Yakut et al. 2004, (179) Tas & Evren 2006, (180) Hambálek & Pribulla 2017, (181) Yang et al. 2005a, (182) Goecking et al. 1994, (183) Kim et al. 2003a, (184) Gomez-Forrellad et al. 1999, (185) Gazeas & Niarchos 2005, (186) Rucinski & Duerbeck 2006, (187) Alvarez et al. 2015, (188) Zhai & Lu 1988, (189) Pribulla2002a, (190) Leung et al. 1985, (191) Kalomeni et al. 2007, (192) Samec & Hube 1991, (193) Lee et al. 2004, (194) Li et al. 2015a, (195) Albayrak et al. 2005, (196) Ekmekçi et al. 2012, (197) Gürol et al. 2011b, (198) Lee et al. 2014, (199) Pi et al. 2017, (200) Barnes et al. 2004, (201) He et al. 2009, (202) Zhang et al. 2020a, (203) Verrot & van Cauteren 2000, (204) Yang et al. 2013, (205) Ulaş & Ulusoy 2014, (206) Sarotsakulchai et al. 2018, (207) Hrivnak et al. 1995, (208) Yue et al. 2019, (209) Bell & Malcolm 1987, (210) Alton et al. 2018, (211) Yang & Liu 2003c,(212) Hrivnak et al. 2006, (213) Li et al. 2014b, (214) Hasanzadeh et al. 2015, (215) Yang & Liu 2003b, (216) Kallrath et al. 2006, (217) Li et al. 2015, (218) Hawkins et al. 2005, (219) Hilditch & King 1986, (220) Chochol et al. 2001, (221) Wang & Lu 1990, (222) Lee et al. 2011, (223) Pribulla et al. 2003b, (224) Wang et al. 2015, (225) Yılmaz et al. 2015, (226) Csák et al. 2000, (227) Selam et al. 2005, (228) Kjurkchieva & Marchev 2010, (229) Zhou et al. 2016b, (230) Kang et al. 2002, (231) Li et al. 2015b, (232) Lu & Rucinski 1993, (233) Hilditch & King 1988, (234) Qian & Yang 2004, (235) Djurašević et al. 2004, (236) Zhu et al. 2013, (237) Kjurkchieva et al. 2019b, (238) Lu et al. 2007, (239) Gürol et al. 2011a, (240) Devarapalli et al. 2020. (This table is available in its entirety in machine-readable form in the online journal. A portion is shown here for guidance regarding its form and content.)

## 3.2. Estimation on the absolute parameters

Based on our best-fitting solutions and *Gaia* distances, we estimate the absolute parameters of the six targets using the following steps. (1) The absolute magnitude, $M_V$, can be calculated by using the following equation, $M_V = m_{V max} - 5 \log D + 5 - A_V$, where $m_{V max}$ is the V band magnitude at maximum which is listed in Table 3, $D$ refers to the *Gaia* distance determined by Bailer-Jones et al. (2018), and $A_V$ is the extinction which can be obtained using the S & F method from the IRAS database. (2) The total luminosity, $L_T$, can be obtained using the following relations, $-2.5 \log L_T/L_\odot = M_{bol} - 4.75$ and $M_{bol} = M_V + BC_V$, where $M_{bol}$ is the absolute bolometric magnitude, while $BC_V$ represents the bolometric correction. (3) Due to the light-curve solutions, the luminosity ratio, $l_2/l_1$, has been determined, which allows us to calculate the luminosities of the two components. (4) The radii, $R_i$, of the two components and the semimajor axis, $a$, can be estimated based on their corresponding luminosities and adopting blackbody emission ($L = 4\sigma R^2 T^4$). (5) The total mass, $M_T$, can be computed based the Kepler's third law $M_T = 0.0134 a^3/P^2$, then the component masses $M_i$ can be derived due to the total mass and the mass ratio $q$. The results are shown in Table 8. The errors of these parameters were determined by error propagation.

## 4. The analysis of the O-C diagram

Continuous orbital period changes, decreasing or increasing, have been discovered for a great number of contact binaries. In order to analyze the continuous period changes of our targets, we collected almost all eclipsing times of them and show them in Table 9 (the eclipsing times were transferred to BJD). All the minima were derived by PE or CCD detectors. Because of the 30-minute cadence of the TESS data, we used the similar method as Wang et al. (2017) and Li et al. (2020) to calculate the eclipsing times. Each sector of the TESS data was divided into four segments, using a linear ephemeris $BJD = BJD_0 + P * E$ (BJD is the observing time, $BJD_0$ is the reference time, P is the orbital period, and E is the cycle number), we shifted the data of one segment into one period, and then we calculated the eclipsing times by using the K-W method (Kwee & van Woerden 1956). Figure 5 displays

Table 8: The fundamental parameters of the six targets

| Star | $D_{Gaia}$ | $A_V$ | $M_V$ | $BC_V$ | $M_{bol}$ | $L_1$ | $L_2$ | $R_1$ | $R_2$ | $a$ | $M_1$ | $M_2$ |
|---|---|---|---|---|---|---|---|---|---|---|---|---|
| | pc | mag | mag | mag | mag | $L_\odot$ | $L_\odot$ | $R_\odot$ | $R_\odot$ | $R_\odot$ | $M_\odot$ | $M_\odot$ |
| V429 Cam | 755.1 ± 9.4 | 0.353 | 3.187 ± 0.027 | -0.048 | 3.139 ± 0.027 | 3.56 ± 0.09 | 0.85 ± 0.02 | 1.55 ± 0.03 | 0.78 ± 0.02 | 2.88 ± 0.08 | 1.36 ± 0.12 | 0.28 ± 0.03 |
| V473 Cam | 208.2 ± 1.1 | 0.116 | 4.690 ± 0.011 | -0.219 | 4.471 ± 0.011 | 0.43 ± 0.01 | 0.84 ± 0.01 | 0.69 ± 0.01 | 1.10 ± 0.02 | 2.26 ± 0.05 | 0.50 ± 0.03 | 1.24 ± 0.09 |
| V505 Cam | 797.0 ± 11.5 | 0.072 | 4.221 ± 0.031 | -0.129 | 4.092 ± 0.032 | 0.60 ± 0.02 | 1.23 ± 0.04 | 0.74 ± 0.02 | 1.15 ± 0.04 | 2.42 ± 0.09 | 0.45 ± 0.06 | 1.23 ± 0.16 |
| V514 Cam | 372.7 ± 3.1 | 0.159 | 4.233 ± 0.018 | -0.131 | 4.102 ± 0.018 | 0.53 ± 0.01 | 1.30 ± 0.02 | 0.71 ± 0.02 | 1.19 ± 0.03 | 2.44 ± 0.08 | 0.35 ± 0.04 | 1.13 ± 0.13 |
| V830 Cep | 348.2 ± 1.8 | 1.116 | 4.575 ± 0.011 | -0.080 | 4.495 ± 0.011 | 0.98 ± 0.01 | 0.29 ± 0.01 | 0.91 ± 0.01 | 0.47 ± 0.01 | 1.74 ± 0.04 | 0.84 ± 0.05 | 0.19 ± 0.01 |
| V1033 Her | 236.8 ± 1.5 | 0.187 | 4.675 ± 0.014 | -0.196 | 4.479 ± 0.014 | 0.39 ± 0.01 | 0.89 ± 0.01 | 0.61 ± 0.02 | 1.08 ± 0.03 | 2.19 ± 0.06 | 0.32 ± 0.03 | 1.26 ± 0.12 |

– 15 –

how to divide one sector data of TESS into four segments and the result of shifting one segment data into one period.

Using the following linear ephemeris,

$$BJD = BJD_0 + P \times E, \tag{2}$$

we calculated the $O - C$ values of our targets. In this equation, the initial epoch, $BJD_0$ ($BJD_0$ is transferred from $HJD_0$), and the orbital period, $P$, were taken from Table 1. Figure 6 plots the $O - C$ diagrams of all the targets. We can identify that five of these targets (except V1033 Her) exhibit secular changes in the $O - C$ diagrams. Then, the following quadratic equation,

$$O - C = \Delta T_0 + \Delta P_0 \times E + \frac{\beta}{2} \times E^2, \tag{3}$$

was used to model their $O - C$ diagrams. For V1033 Her, there is a periodic variation in addition to the secular change, so we used the following equation to fit its $O - C$ diagram,

$$O - C = \Delta T_0 + \Delta P_0 \times E + \frac{\beta}{2} \times E^2 + A \times \sin(\frac{2\pi}{P_3} \times E + \varphi) \tag{4}$$

The corresponding results of Equations (3) and (4) are exhibited in Table 10. In order to compare the goodness between the linear and quadratic fits of the $O - C$ diagrams for five of our six targets (except V1033 Her, we compared the goodness between the quadratic and quadratic plus sinusoidal fits for V1033 Her), the Bayes information criterion ($BIC$) method (e.g., Burnham & Anderson 2002) was adopted. The $BIC$ method was defined as follows,

$$BIC = n \log(RSS/n) + k \log n, \tag{5}$$

where $RSS$ is the residual sum of squares by the linear or quadratic fitting, $n$ is the data number, while $k$ is the number of the fitted parameters. The values of $RSS$ and $BIC$ of the linear and quadratic fits (the quadratic and quadratic plus sinusoidal fits for V1033 Her) are shown in Table 10. Statistically, the quadratic fit is better than the linear fit for four targets (except V473 Cam), and the quadratic plus sinusoidal fit is better than the quadratic fit for V1033 Her. Therefore, only the linear part of Equation (3) was applied to V473 Cam to analyze its $O - C$ diagram. Due to our analysis, two binaries manifest long-term period increase, while three show secular period decrease. For V1033 Her, there is a cyclic variation with a period of $8.61 \pm 0.23$ yr and an amplitude of $0.0021 \pm 0.0002$ days superimposed on the long-term period increase. If these long-term period changes are due to mass transfer, we can compute the mass transfer rate by using the equation below,

$$\frac{\dot{P}}{P} = -3\dot{M}_1(\frac{1}{M_1} - \frac{1}{M_2}). \tag{6}$$

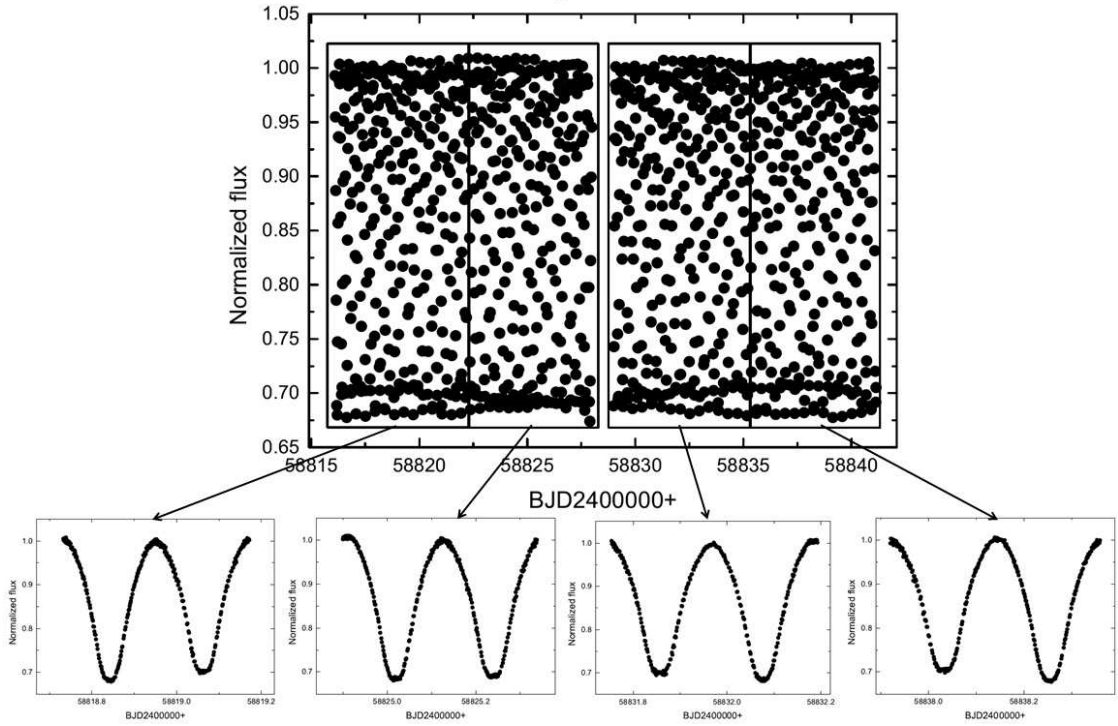

Fig. 5.— The upper panel shows the TESS data of V429 Cam, we divided them into four segments and showed in the four black rectangles. Then, using the equation $BJD = BJD_0 + P * E$ (BJD is the observing time, $BJD_0$ is the reference time, P is the orbital period, and E is the cycle number), we shifted one segment data into one period and displayed them in the lower panel.

The results are also shown in Table 10. Actually, the reliability of the $O - C$ variations of V429 Cam, V505 Cam, and V830 Cep should be treated with caution due to the insufficient times of minima. Future observations are needed to check our results.

## 5. Discussions and conclusions

To study the formation, evolution, structure, and final destiny of contact binaries, absolute parameters should be derived for as many contact binaries as possible. By collecting 173 contact binaries whose absolute parameters have been obtained, we found that the distances calculated by absolute parameters are agree with their *Gaia* distances within an accuracy of 25% for more than eighty percent of these contact binaries. Therefore, *Gaia* distance gives researchers an opportunity on estimating absolute parameters for most contact binaries without radial velocity curves. Six totally eclipsing contact binaries with symmetric light curves were observed and analyzed. We found that only one target is an A-type contact binary (V429 Cam), while the others are W-type contact ones. Two of the six targets are medium contact systems, while the others are shallow contact ones. In Table 7, both spectroscopic and photometric mass ratios of 101 contact binaries have been determined. Figure 7 plots the relationship between the spectroscopic and photometric mass ratios. In the 101 systems, 59 are total eclipse binaries. For about half of the partial eclipse systems, their photometric mass ratios are very different from the spectroscopic ones, while for almost all the total eclipse systems, their photometric mass ratios are in good agreement with the spectroscopic ones, indicating that the photometric mass ratios derived for totally eclipsing contact binaries are reliable. This result is consistent with those determined by Pribulla et al. (2003a) and Terrell & Wilson (2005). Therefore, the light-curve solutions for the six targets should be accurate and reliable. According to the light-curve solutions and *Gaia* distances, we estimated the absolute parameters for these six targets. By collecting the times of light minima, we investigated the period variations of them and discovered that two of them exhibit secular period increase while three manifest long-term period decrease.

Based on the absolute parameters listed in Tables 7 and 8, we discussed the evolutionary states of the contact binaries. These contact binaries were divided into two types, A-type and W-type (94 A-type systems, while 85 W-type ones). Figure 8 plots the relations of Mass-Radius (M-R) and Mass-Luminosity (M-L). The solid and dashed lines represent the zero-age main-sequence (ZAMS) and the terminal-age main-sequence (TAMS) which are constructed by the binary star evolution code provide by Hurley et al. (2002), the solid and open circles refer to the more massive components and the less massive ones of A-type contact binaries, while the solid and open triangles denote the more massive components and the less massive

Table 9: The eclipsing times of our six targets

| Star | BJD | Error | E | O-C | Res. | Ref. |
|------|-----|-------|---|-----|------|------|
| | 2400000+ | | | | | |
| V429 Cam | 51515.0419 | — | -16598 | 0.0509 | 0.00004 | (1) |
| | 55970.4834 | 0.0017 | -6498.5 | -0.0062 | 0.0010 | (2) |
| | 56535.3859 | 0.0013 | -5218 | -0.0094 | -0.0011 | (3) |
| | 58818.8421 | 0.0002 | -42 | 0.0011 | 0.0003 | (4) |
| | 58819.0627 | 0.0003 | -41.5 | 0.0011 | 0.0003 | (4) |
| | 58825.0185 | 0.0002 | -28 | 0.0012 | 0.0004 | (4) |
| | 58825.2386 | 0.0003 | -27.5 | 0.0008 | -0.0001 | (4) |
| | 58831.8569 | 0.0003 | -12.5 | 0.0017 | 0.0008 | (4) |
| | 58832.0765 | 0.0002 | -12 | 0.0007 | -0.0002 | (4) |
| | 58837.3697 | 0.0003 | 0 | 0.0000 | -0.0010 | (5) |
| | 58838.0328 | 0.0003 | 1.5 | 0.0013 | 0.0003 | (4) |
| | 58838.2536 | 0.0002 | 2 | 0.0016 | 0.0006 | (4) |
| | 58842.0017 | 0.0005 | 10.5 | -0.0002 | -0.0012 | (5) |

(1) Khruslov 2006; (2) IBVS 6048; (3) IBVS 6118; (4) This paper (TESS); (5) This paper; (6) IBVS 5918; (7) IBVS 6029; (8) IBVS 6070; (9) IBVS 6131; (10) Kjurkchieva et al. 2017; (11) IBVS 6196; (12) http://var2.astro.cz/brno/protokoly.php; (13) IBVS 6234; (14) Khruslov 2008; (15) IBVS 6026; (16) IBVS 6149; (17) IBVS 6157; (18) IBVS 6244; (19) BAV 031; (20) Kuzmin 2007; (21) IBVS 5060; (22) BBSAG 126; (23) BBSAG 128; (24) https://www.aavso.org/bob-nelsons-o-c-fles; (25) IBVS 5438; (26) This paper (Super-WASP); (27) IBVS 5653; (28) IBVS 5731; (29) IBVS 5713; (30) IBVS 5802; (31) IBVS 5781; (32) IBVS 5874; (33) IBVS 5871; (34) IBVS 5929; (35) IBVS 5920; (36) IBVS 5945; (37) IBVS 5959; (38) IBVS 6010; (39) IBVS 5992; (40) IBVS 6041; (41) JAAVSO 44-1; (42) JAAVSO 45-1; (43) JAAVSO 46-2; (44) BRNO 41; (45) Long et al. 2019.

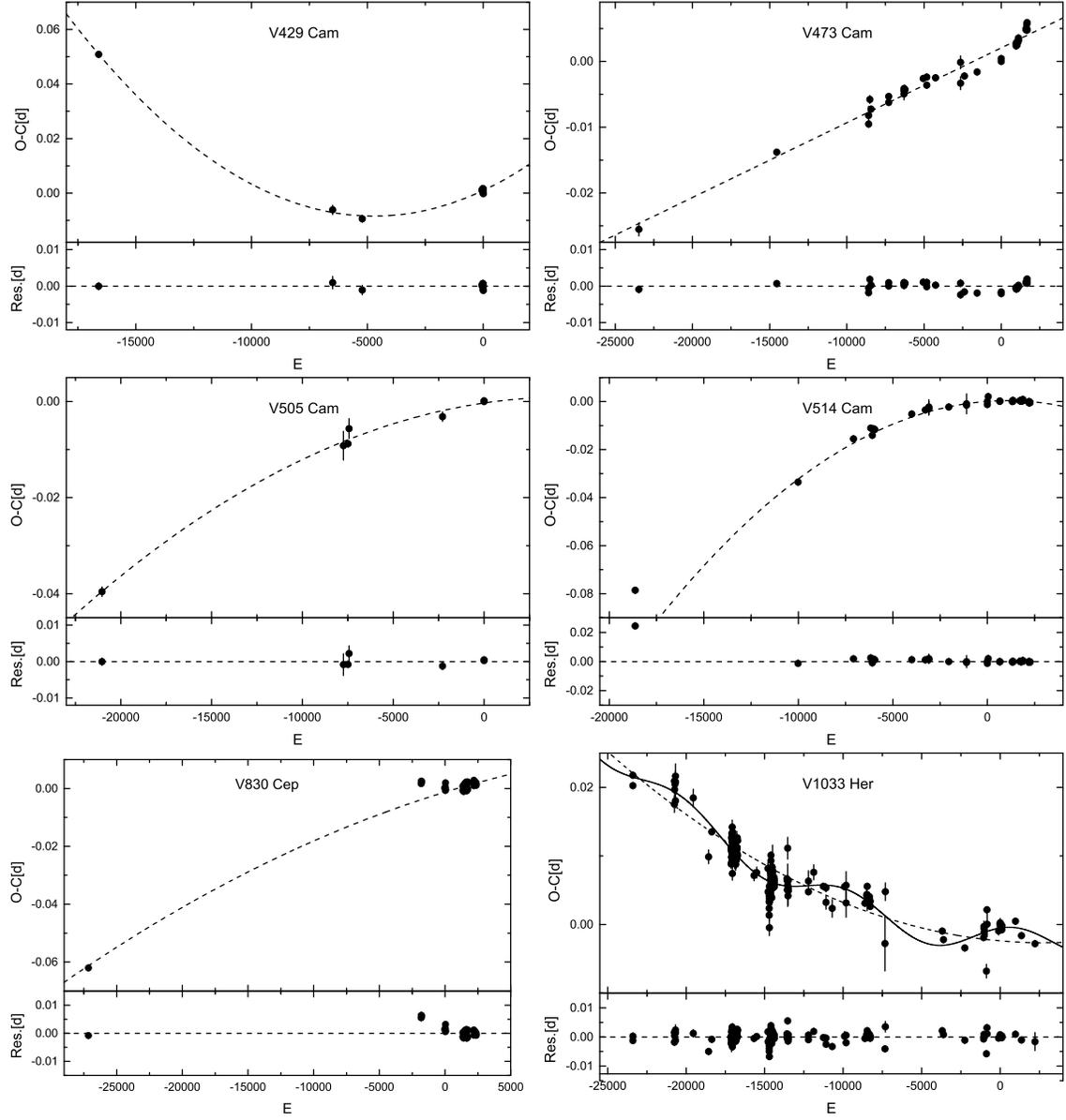

Fig. 6.— The $O - C$ diagrams for our six targets. The errors were set as 0.001 for the minima that are not given errors in the literature.

Table 10: The parameters determined by Equations (3), (4), (5), and (6) for our six targets

| Star | $\Delta T_0$ ($\times 10^{-4}$ d) | $\Delta P_0$ ($\times 10^{-7}$ d) | $\beta$ ($\times 10^{-7}$ d yr$^{-1}$) | $dM_1/dt$ ($\times 10^{-7}$ $M_\odot$ yr$^{-1}$) | $RSS_L$ ($\times 10^{-6}$) | $RSS_Q$ ($\times 10^{-6}$) | $BIC_L$ | $BIC_Q$ |
|---|---|---|---|---|---|---|---|---|
| V429 Cam | $9.68 \pm 2.42$ | $39.81 \pm 1.56$ | $6.97 \pm 0.17$ | $1.86 \pm 0.04$ | 1042 | 6 | -51 | -79 |
| V473 Cam | $20.5 \pm 1.74$ | $11.38 \pm 3.02$ | – | – | 48 | 48 | -272 | -270 |
| V505 Cam | $-2.83 \pm 9.31$ | $5.54 \pm 2.50$ | $-1.36 \pm 0.25$ | $9.53 \pm 0.18$ | 69 | 8 | -33 | -39 |
| V514 Cam | $0.84 \pm 2.31$ | $5.62 \pm 1.03$ | $-5.38 \pm 0.28$ | $1.73 \pm 0.14$ | 1690 | 630 | -210 | -229 |
| V830 Cep | $-12.4 \pm 5.11$ | $13.9 \pm 2.65$ | $-0.85 \pm 0.54$ | $-0.27 \pm 0.17$ | 692 | 283 | -314 | -337 |
| V1033 Her | $-23.7 \pm 4.49$ | $-1.95 \pm 0.85$ | $0.89 \pm 0.09$ | $-0.43 \pm 0.04$ | 729[a] | 528[b] | -1299[a] | -1330[b] |

[a] These results were derived based on the quadratic fitting. [b] These results were derived based on the quadratic plus sinusoidal fitting.

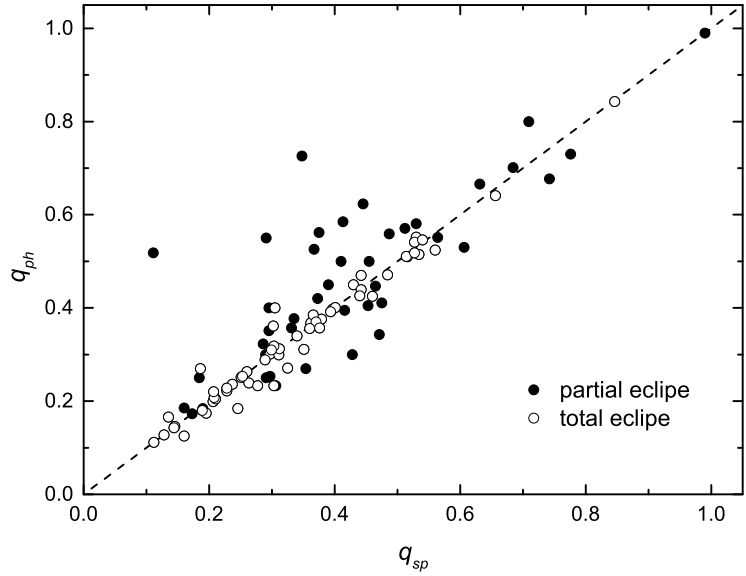

Fig. 7.— The relation between the spectroscopic and photometric mass ratios.

ones of W-type contact binaries, and the red symbols show the two components of our six targets. As seen in this figure, the more massive primaries are fairly close to the ZAMS comparing to the less massive secondaries, meaning that the more massive primaries are less evolved than the less massive secondaries, and the less massive secondaries are over-sized and over-luminous in terms of their masses. This may be resulted from the past energy and mass transfer from the primary to the secondary. These results are similar with those determined by previous researches (e.g., Yakut & Eggleton 2005). We also found that for the same mass, the luminosity and radius of each component of A-type contact binaries are greater than those of W-type contact systems, indicating that A-type contact binaries are more evolved than W-type contact systems (e.g., Lucy 1976; Wilson 1978). This may mean that the two subtype contact binaries have different formation channels (e.g., Zhang et al. 2020b)

The orbital angular momentums $J_{orb}$ of the contact binaries were calculated by using the following equation (Christopoulou & Papageorgiou 2013),

$$J_{orb} = 1.24 \times 10^{52} \times M_T^{5/3} \times P^{1/3} \times q \times (1+q)^{-2}, \tag{7}$$

where $M_T$ is the total mass of a binary in solar units, $P$ represents the orbital period in days, and $q$ is the mass ratio. The obtained results are displayed in Figure 9. The detached binaries and the boundary line separating detached and contact systems taken from Eker et al. (2006) are also plotted in this figure. We found that almost all the contact binaries are below the boundary line, indicating that the $J_{orb}$ of contact binaries are smaller than those of detached binaries with the same masses. This discrepancy is caused by the past angular momentum loss during the formation and evolution of contact binaries due to orbital period decrease or/and mass loss. This is a proof that the contact binaries are formed from short period detached binaries by angular momentum loss. In addition, we found that the A-type systems have lower angular momenta than W-type systems, indicating that they are more evolved, this is corresponding to the result derived by the M-R and M-L relations.

In conclusion, six totally contact binaries with symmetric light curves were observed and analyzed. By comparing the *Gaia* distances with those computed by fundamental parameters of contact binaries, we found that *Gaia* distances can be used to estimate absolute parameters for most contact binaries. We applied the *Gaia* distances to estimate the absolute parameters of our six systems. Based on our collections and the six targets, the evolutionary states of contact binaries were analyzed. *Gaia* distances provide the opportunity to estimate the global parameters for a great number of contact binaries, especially those identified by the worldwide photometric surveys because of the public available photometric light curves. These results can be applied to improve the theoretical models of the formation, evolution, structure, and ultimate fate of contact binaries.

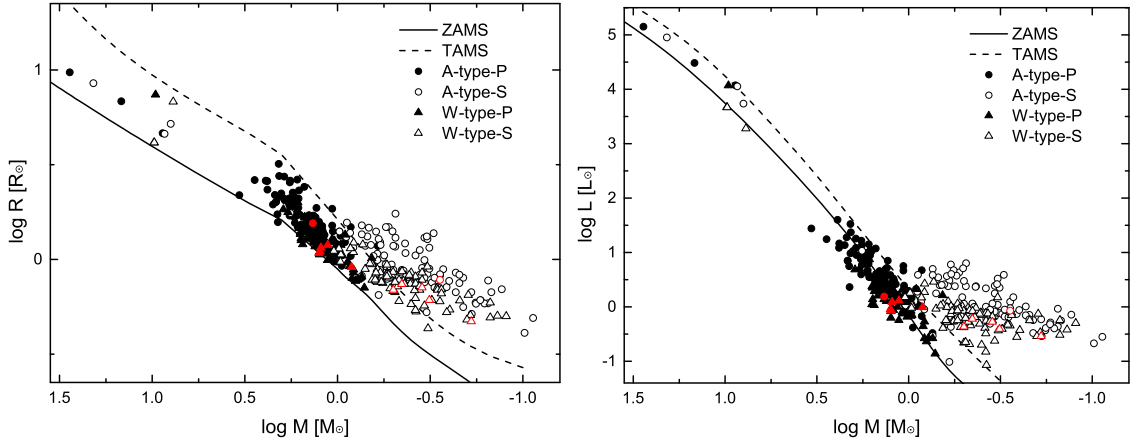

Fig. 8.— The relations of M-R and M-L for contact binaries. The solid and dashed lines represent the ZAMS and the TAMS which are constructed by the binary star evolution code provide by Hurley et al. (2002), the solid and open circles refer to the more massive components and the less massive ones of A-type contact binaries, while the solid and open triangles denote the more massive components and the less massive ones of W-type contact binaries, and the red symbols show the two components of our six targets.

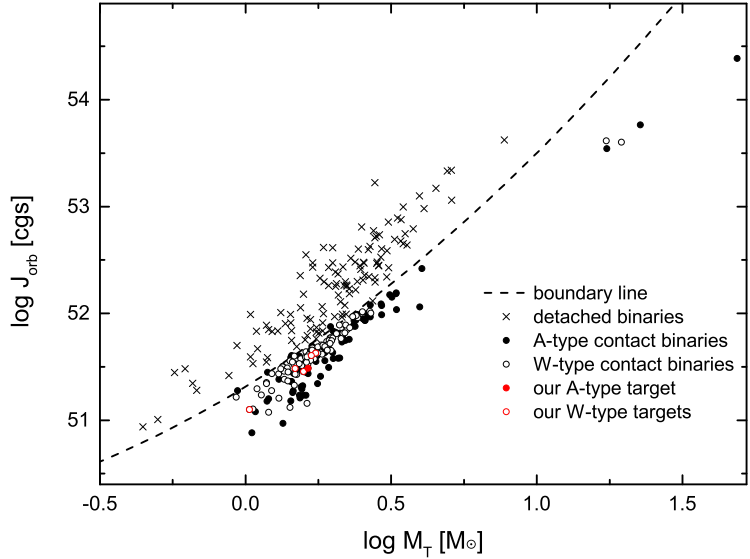

Fig. 9.— The relation between orbital angular momentum and total mass for detached and contact binaries, the detached binaries and the boundary line separating detached and contact systems were taken from Eker et al. (2006). The contact binaries were those listed in Tables 6 and 7. The black crosses denote the detached binaries, the black solid and open circles refer to the A-type and W-type contact binaries, respectively, while the red symbols represent our six targets.

## 6. Acknowledgments

We thank the anonymous referee very much for the valuable suggestions and comments which have improved this manuscript greatly. This work is supported by the Joint Research Fund in Astronomy (No. U1931103) under cooperative agreement between National Natural Science Foundation of China (NSFC) and Chinese Academy of Sciences (CAS), and by NSFC (No. 11703016), and by the Natural Science Foundation of Shandong Province (Nos. ZR2014AQ019, JQ201702), and by Young Scholars Program of Shandong University, Weihai (Nos. 20820162003, 20820171006), and by the Chinese Academy of Sciences Interdisciplinary Innovation Team, and by the Open Research Program of Key Laboratory for the Structure and Evolution of Celestial Objects (No. OP201704). Work by C.H.K. was supported by the grant of National Research Foundation of Korea (2020R1A4A2002885).

This work is partly supported by the Supercomputing Center of Shandong University, Weihai.

The spectral data were provided by Guoshoujing Telescope (the Large Sky Area Multi-Object Fiber Spectroscopic Telescope LAMOST), which is a National Major Scientific Project built by the Chinese Academy of Sciences. Funding for the project has been provided by the National Development and Reform Commission. LAMOST is operated and managed by the National Astronomical Observatories, Chinese Academy of Sciences.

We acknowledge the support of the staff of the Xinglong 60cm telescope, NEXT and WHOT. This work was partially supported by the Open Project Program of the Key Laboratory of Optical Astronomy, National Astronomical Observatories, Chinese Academy of Sciences.

This work has made use of data from the European Space Agency (ESA) mission *Gaia* (https://www.cosmos.esa.int/gaia), processed by the *Gaia* Data Processing and Analysis Consortium (DPAC, https://www.cosmos.esa.int/web/gaia/dpac/consortium). Funding for the DPAC has been provided by national institutions, in particular the institutions participating in the *Gaia* Multilateral Agreement.

This work includes data collected by the TESS mission. Funding for the TESS mission is provided by NASA Science Mission directorate. We acknowledge the TESS team for its support of this work.

This publication makes use of data products from the AAVSO Photometric All Sky Survey (APASS). Funded by the Robert Martin Ayers Sciences Fund and the National Science Foundation.

This publication makes use of data products from the Two Micron All Sky Survey,

which is a joint project of the University of Massachusetts and the Infrared Processing and Analysis Center/California Institute of Technology, funded by the National Aeronautics and Space Administration and the National Science Foundation.

This paper makes use of data from the DR1 of the WASP data (Butters et al. 2010) as provided by the WASP consortium, and the computing and storage facilities at the CERIT Scientific Cloud, reg. no. CZ.1.05/3.2.00/08.0144 which is operated by Masaryk University, Czech Republic.

Funding for SDSS-III has been provided by the Alfred P. Sloan Foundation, the Participating Institutions, the National Science Foundation, and the U.S. Department of Energy Office of Science. The SDSS-III web site is http://www.sdss3.org/.

SDSS-III is managed by the Astrophysical Research Consortium for the Participating Institutions of the SDSS-III Collaboration including the University of Arizona, the Brazilian Participation Group, Brookhaven National Laboratory, University of Cambridge, University of Florida, the French Participation Group, the German Participation Group, the Instituto de Astrofisica de Canarias, the Michigan State/Notre Dame/JINA Participation Group, Johns Hopkins University, Lawrence Berkeley National Laboratory, Max Planck Institute for Astrophysics, New Mexico State University, New York University, Ohio State University, Pennsylvania State University, University of Portsmouth, Princeton University, the Spanish Participation Group, University of Tokyo, University of Utah, Vanderbilt University, University of Virginia, University of Washington, and Yale University.

# REFERENCES

Akalin, A., & Derman, E. 1997, A&AS, 125, 407

Albayrak, B., Djurašević, G., Selam, S. O., Atanacković-Vukmanović, O., & Yılmaz, M. 2005, NewA, 10, 163

Alton, K. B., & Nelson, R. H. 2018, MNRAS, 479, 3197

Alton, K. B., Nelson, R. H., & Terrell, D. 2018, IBVS, 6256, 1

Alton, K. B., & Terrell, D. 2006, JAVSO, 34, 188

Alvarez, G. E., Sowell, J. R., Williamon, R. M., & Lapasset, E. 2015, PASP, 127, 742

Arbutina, B. 2007, MNRAS, 377, 1635

Bailer-Jones, C. A. L., Rybizki, J., Fouesneau, M., Mantelet, G., & Andrae, R. 2018, AJ, 156, 58

Baran, A., Zola, S., Rucinski, S. M., Kreiner, J. M., Siwak, M., & Drodz, M. 2004, AcA, 54, 195

Barden, S. C. 1987, ApJ, 317, 333

Barnes, J. R., Lister, T. A., Hilditch, R. W., & Collier Cameron, A. 2004, MNRAS, 348, 1321

Barone, F., di Fiore, L., Milano, L., & Russo, G. 1993, ApJ, 407, 237

Bell, S. A., & Malcolm, G. J. 1987, MNRAS, 226, 899

Binnendijk, L. 1970, VA, 12, 217

Binnendijk, L. 1984, PASP, 96, 646

Bradstreet, D. H., & Guinan, E. F. 1994, in ASP Conf. Ser. 56, Interacting Binary Stars, ed. Allen W. Shafter (San Francisco, CA: ASP), 228

Butters, O. W., West, R. G., Anderson, D. R. et al. 2010, A&A, 520, 10

Burnham, K. P., & Anderson, D. R. 2002, Model Selection and Multimodel Inference: A Practical Information-Theoretic Approach, 2nd edn. Springer, New York

Çalışkan, Ş., Latković, O., Djurašević, G. et al. 2014, AJ, 148, 126

Caton, D., Gentry, D. R., Samec, R. G. et al. 2019, PASP, 131, 054203

Chochol, D., van Houten, C. J., Pribulla, T., Grygar, J. 2001, CoSka, 31, 5

Christopoulou, P.-E., & Papageorgiou, A. 2013, AJ, 146, 157

Christopoulou, P.-E., Parageorgiou, A., & Chrysopoulos, I. 2011, AJ, 142, 99

Christopoulou, P.-E., Papageorgiou, A., Vasileiadis, T., & Tsantilas, S. 2012, AJ, 144,149

Çiçek, C. 2011, NewA, 16, 12

Covey, K., Ivezić, Ž., Schlegel, D., et al. 2007, AJ, 134, 2398

Csák, B., Kiss, L. L., Vinkó, J., Alfaro, E. J 2000, A&A, 356, 603

Cutri, R. M., Skrutskie, M. F., van Dyk, S., et al. 2003, VizieR Online Data Catalog, 2246, 0

Deb, S. & Singh, H. P. 2011, MNRAS, 412, 1787

Degirmenci, O. L. 2006, IBVS, 5726, 1

Degirmenci, O. L., Sezer, C., Demircan, O., Erdem, A., Özdemir, S., Ak, H., & Albayrak, B. 1999, A&AS, 134, 327

Devarapalli, S. P., Jagirdar, R., Prasad, R. M. et al. 2020, MNRAS, 495, 1565

Djurašević, G., Albayrak, B., Tanrıverdi, T., & Erkapić, S. 2004, A&A, 415, 283

Djurašević, G., Baştürk, Ö., Latković, O. et al. 2013, AJ, 145,80

Djurašević, G., Dimitrov, D., Arbutina, B., Albayrak, B., Selam, S. O., & Atanacković-Vukmanović, O. 2006, PASA, 23, 154

Djurašević, G., Yılmaz, M., Baştürk, Ö., Kılıçoğlu, T., Latković, O., & Çalışkan, Ş. 2011, A&A, 525, 66

Duerbeck, H. W., & Rucinski, S. M. 2007, AJ, 133, 169

Dumitrescu, A. 2003, RoAJ, 13, 119

Drechsel, H., Rahe, J., Wargau, W., & Wolf, B. 1982, A&A, 110, 246

Eggleton, P. P. 2012, JASS, 29, 145

Eggleton, P. P., & Kiseleva-Eggleton, L. 2002, ApJ, 575, 461

Eker, Z., Demircan, O., Bilir, S., & Karataş, Y. 2006, MNRAS, 373, 1483

Ekmekçi, F., Elmaslı, A., Yılmaz, M. et al. 2012, NewA, 17, 603

Erdem, A., & Özkardeş, B. 2006, NewA, 12, 192

Erdem, A., & Özkardeş, B. 2009, NewA, 14, 321

Erkan, N., & Ulaş, B. 2016, NewA, 46, 73

Essam, A., Saad, S. M., Nouh, M. I., Dumitrescu, A., El-Khateeb, M. M., & Haroon, A. 2010, NewA, 15, 227

Flannery, B. P. 1976, ApJ, 205, 217

Gaia Collaboration, Prusti, T., de Bruijne, J. H. J. et al. 2016, A&A, 595, 1

Gaia Collaboration, Brown, A. G. A., Vallenari, A. et al. 2018, A&A, 616, 1

Gazeas, K. D., Baran, A., Niarchos, P. et al. 2005, AcA, 55, 123

Gazeas, K. D., & Niarchos, P. G. 2005, ARBl, 20, 193

Gazeas, K. D., Niarchos, P. G., Zola, S., Kreiner, J. M., & Rucinski, S. M. 2006, AcA, 56, 127

Goecking, K.-D., & Duerbeck, H. W. 1993, A&A, 278, 463

Goecking, K.-D., Duerbeck, H. W., Plewa, T. et al. 1994, A&A, 289, 827

Gomez-Forrellad, J. M., Garcia-Melendo, E., Guarro-Flo, J., Nomen-Torres, J., & Vidal-Sainz, J. 1999, IBVS, 4702, 1

Gorda, S. Y. 2016, AstBu, 71, 64

Gu, S.-H. 1999, A&A, 346, 437

Gürol, B., Bradstreet, D. H., Demircan, Y., & Gürsoytrak, S. H. 2015a, NewA, 41, 26

Gürol, B., Bradstreet, D. H., & Okan, A. 2015b, NewA, 36, 100

Gürol, B., Derman, E., Saguner, T. et al. 2011a, NewA, 16, 242

Gürol, B., Gökay, G., Saral, G., Gürsoytrak, S. H., Cerit, S., & Terzioğlu, Z. 2016, NewA, 46, 31

Gürol, B., Gürsoytrak, S. H., & Bradstreet, D. H. 2015c, NewA, 39, 9

Gürol, B., Terzioğlu, Z., Gürsoytrak, S. H., Gökay, G., & Derman, E. 2011b, AN, 332, 690

Hambálek, L., & Pribulla, T. 2017, ASPC, 510, 372

Hasanzadeh, A., Farsian, F., & Nemati, M. 2015, NewA, 34, 262

Harries, T. J., Hilditch, R. W., & Hill, G. 1997, MNRAS, 285, 277

Hawkins, N. C., Samec, R. G., Van Hamme, W., & Faulkner, D. R. 2005, IBVS, 5612, 1

He, J.-J., & Qian, S.-B. 2008, ChJAA, 8, 465

He, J.-J., Qian, S.-B., Fernández Lajús, E., & Fariña, C. 2009, AJ, 138, 1465

Henden, A. A., Levine, S., Terrell, D., & Welch, D. L. 2016, VizieR Online Data Catalog, 2336, 0

Hilditch, R. W. 1981, MNRAS, 196, 305

Hilditch, R. W., Hill, G., & Bell, S. A. 1992, MNRAS, 255, 285

Hilditch, R. W., & King, D. J. 1986, MNRAS, 223, 581

Hilditch, R. W., & King, D. J. 1988, MNRAS, 231, 397

Hilditch, R. W., King, D. J., Hill, G., & Poeckert, R. 1984, MNRAS, 208, 135

Hilditch, R. W., King, D. J., & McFarlane, T. M. 1989, MNRAS, 237, 447

Hill, G., Fisher, W. A., & Holmgren, D. 1989, A&A, 211, 81

Hiller, M. E., Osborn, W., & Terrell, D. 2004, PASP, 116, 337

Hoffman, D. I., Harrison, T. E., & McNamara, B. J. 2009, AJ, 138, 466

Hoffmann, M. 1978, A&AS, 33, 63

Hu, S. M., Han, S. H., Guo, D. F., & Du, J. J. 2014, RAA, 14, 719

Hurley, J. R., Tout, C. A., & Pols, O. R. 2002, MNRAS, 329, 897

Hrivnak, B. J. 1988, ApJ, 335,319

Hrivnak, B. J. 1989, ApJ, 340, 458

Hrivnak, B. J. 1993, ASPC, 38, 269

Hrivnak, B. J., Guinan, E.F., & Lu, W.-X. 1995, ApJ, 455, 300

Hrivnak, B. J., Lu, W.-X. Eaton, J., & Kenning, D. 2006, AJ, 132, 960

Jayasinghe, T., Kochanek, C. S., Stanek, K. Z. et al. 2018, MNRAS, 477, 3145

Jiang, D., Han, Z., Wang, J., Jiang, T., & Li, L. 2010, MNRAS, 405, 2485

Jiang, D., Han, Z., Ge, H., Yang, L., & Li, L. 2012, MNRAS, 421, 2769

Kallrath, J., Milone, E. F., Breinhorst, R. A., Wilson, R. E., Schnell, A., & Purgathofer, A. 2006, A&A, 452, 959

Kalomeni, B., Yakut, K., Keskin, V. et al. 2007, AJ, 134, 642

Kaluzny, J. 1984, AcA, 34, 217

Kaluzny 1986, AcA, 36, 113

Kaluzny, J., & Rucinski, S. M. 1986, AJ, 92, 666

Kang, Y.-W., Oh, K.-D., Kim, C.-H., Hwang, C., Kim, H.-I., & Lee, W.-B. 2002, MNRAS, 331, 707

Khruslov, A. V. 2006, PZP, 6, 11

Khruslov, A. V. 2008 PZP, 8, 40

Kim, C.-H., Lee, J.-W., Kim, H.-I., Kyung, J.-M., & Koch, R. H. 2003a, AJ, 126, 1555

Kim, C.-H., Lee, J.-W., Kim, S.-L., Han, W., & Koch, R. H. 2003b, AJ, 125, 322

King, D. J., & Hilditch, R. W. 1984, MNRAS, 209, 645

Kjurkchieva, D. P., & Marchev, D. V. 2010, ASPC, 435, 111

Kjurkchieva, D. P., Marchev, D. V., & Popov, V. A. 2018, AN, 339, 472

Kjurkchieva, D. P., Popov, V. A., & Petrov, N. I. 2019a, AJ, 158, 186

Kjurkchieva, D. P., Popov, V. A., & Petrov, N. I. 2020, NewA, 77, 101325

Kjurkchieva, D. P., Popov, V. A., Vasileva, D. L., & Petrov, N. I. 2017, RMxAA, 53, 235

Kjurkchieva, D. P., Stateva, I., Popov, V. A., & Marchev, D. 2019b, AJ, 157, 73

Köse, O., Kalomeni, B., Keskin, V., Ulaş, B., & Yakut, K. 2011, AN, 332, 626

Kreiner, J. M., Rucinski, S. M., Zola, S. et al. 2003, A&A, 412, 465

Krzesiński, J., Mikołajewski, M., Pajdosz, G., & Zola, S. 1991, Ap&SS, 184, 37

Kuzmin, M. L. 2007, PZP, 7, 31

Kwee, K. K. & van Woerden, H. 1956, BAN, 12, 327

Lapasset, E., & Gomez, M. 1990, A&A, 231, 365

Lee, J.-W., Kim, C.-H., Han, W.-Y., Kim, H.-I., & Koch, R. H. 2004, MNRAS, 352, 1041

Lee, J.-W., Kim, S.-L., Lee, Ch.-U., & Youn, J.-H. 2009, PASP, 121, 1366

Lee, J.-W., Lee, C.-U., Kim, S.-L., Kim, H.-I., & Park, J.-H. 2011, PASP, 123, 34

Lee, J.-W., & Park, J.-H. 2018, PASP, 130, 4201

Lee, J.-W., Park, J.-H., Hong, K.-S., Kim, S.-L., & Lee, C.-U. 2014, AJ, 147, 91

Lee, J.-W., Youn, J.-H., Park, J.-H., & Wolf, M. 2015, AJ, 149, 194

Leung, K.-C. 1976, PASP, 88, 936

Leung, K.-C., & Schneider, D. P. 1978, ApJ, 222, 917

Leung, K.-C., Zhai, D., & Zhang, Y. 1985, AJ, 90, 515

Liao, W.-P., Qian, S.-B., & Sarotsakulchai, T. 2019, AJ, 157, 207

Li, L., & Zhang, F. 2006, MNRAS, 369, 2001

Li, K., Gao, D.-Y., Hu, S.-M., Guo, D.-F., Jiang, Y.-G., & Chen, X. 2016, Ap&SS, 361, 63

Li, K., Hu, S. -M., Guo, D.-F., Jiang, Y.-G., Gao, D.-Y., & Chen, X. 2015a, NewA, 41, 17

Li, K., Hu, S. -M., Guo, D.-F., Jiang, Y.-G., Gao, D.-Y., Chen, X., & Odell, A. P. 2015b, AJ, 149, 120

Li, K., Hu, S. -M., Jiang, Y. -G., Chen, X., & Ren, D. -Y. 2014a, NewA, 30, 64

Li, Kai, Kim, Chun-Hwey, Xia, Qi-Qi ey al. 2020, AJ, 159, 189

Li, K., & Qian, S.-B. 2013, NewA, 21, 46

Li, K., Qian, S.-B., Hu, S.-M., & He, J.-J. 2014b, AJ, 147, 98

Li, K., Xia, Q.-Q., Liu, J.-Z. et al. 2019a, RAA, 19, 147

Li, K., Xia, Q.-Q., Michel, R. et al. 2019b, MNRAS, 485, 4588

Liu, Q., Soonthornthum, B., Yang, Y. et al. 1996, A&AS, 118, 453

Liu, J., Esamdin, A., Zhang, Y. et al. 2019, PASP, 131, 4202

Liu, L., Qian, S.-B., He, J.-J., Liao, W.-P., Zhu, L.-Y., & Zhao, E.-G. 2012, PASJ, 64, 48

Liu, L., Qian, S.-B., Zhu, L.-Y., He, J.-J., & Li, L.-J. 2011, AJ, 141,147

Liu, Q., & Yang, Y. 2003, ChJAA, 3, 142

Liu, Q.,, Yang, Y., Gu, C., & Wang, B. 1993, A&AS, 101, 253

Long L., Zhang L.-Y., Han X. L., Lu H.-P., Pi Q.-F., & Yue Q. 2019, MNRAS, 487, 5520

Lorenz, R., Mayer, P., & Drechsel, H. 1999, A&A, 345, 531

Lorenzo, J., Negueruela, I., Vilardell, F. et al. 2016, A&A, 590, 45

Lu, W.-X. 1991, AJ, 102, 262

Lu, W.-X., Hrivnak, B. J., & Rush, B. W. 2007, AJ, 133, 255

Lu, W.-X., & Rucinski, S. M. 1993, AJ, 106, 361

Lu, W.-X., & Rucinski, S. M. 1999, AJ, 118, 515

Lu, W.-X., Rucinski, S. M., & Ogłoza, W. 2001, AJ, 122, 402

Lucy, L. B. 1968, ApJ, 151, 1123

Lucy, L. B. 1976, ApJ, 205, 208

Lucy, L. B., & Wilson, R. E. 1979, ApJ, 231, 502

Luo, A.-L., Zhao, Y.-H., Zhao, G., et al. 2015a, RAA, 15, 1095

Luo, C.-Q., Zhang, X.-B., Deng, L.-C. Wang, K., & Luo, Y.-P. 2015b, AJ, 150, 70

Luo, X., Wang, K., Zhang, X.-B. et al. 2017, AJ, 154, 99

Maceroni, C., Milano, L., & Russo, G. 1982, A&AS, 49, 123

Maceroni, C., Milano, L., & Russo, G. 1984, A&AS, 58, 405

Markworth, N. L., & Michaels, E. J. 1982, PASP, 94, 350

Mauder, H. 1972, A&A, 17, 1

Metcalfe, T. S. 1999, AJ, 117, 2503

McLean, B. J., & Hilditch, R. W. 1983, MNRAS, 203, 1

Milone, E. F., Groisman, G., Fry, D. J. I., Bradstreet, D. H. 1991, ApJ, 370, 677

Milone, E. F., Wilson, R. E., & Hrivnak, B. J. 1987, ApJ, 319, 325

Mitnyan, T., Bódi, A., Szalai, T. et al. 2018, A&A, 612, 91

Molik, P., & Wolf, M. 2000, IBVS, 4951, 1

Nelson, R. H. 2010, IBVS, 5951, 1

Nelson, R. H., & Alton, K. B. 2019, IBVS, 6266, 1

Nelson, R. H., Milone, E. F., van Leeuwen, J., Terrell, D., Penfold, J. E., & Kallrath, J. 1995, AJ, 110, 2400

Nelson, R. H., Şenavci, H. V., Baştürk, Ö., & Bahar, E. 2014, NewA, 29, 57

Nesci, R., Maceroni, C., Milano, L., & Russo, G. 1986, A&A, 159, 142

Niarchos, P. G. 1978, Ap&SS, 58, 301

Niarchos, P. G., Hoffmann, M., & Duerbeck, H. W. 1994a, A&A, 292, 494

Niarchos, P. G., Hoffmann, M., & Duerbeck, H. W. 1994b, A&AS, 103, 39

Nomen-Torres, J., & Garcia-Melendo, E. 1996, IBVS, 4365, 1

Noori, H. R., & Abedi, A. 2017, NewA, 56, 5

O'Connell, D. J. K. 1951, PRCO, 2, 85

Oh, K.-D., Kim, H.-I., & Sung, E.-C. 2010, JASS, 27, 69

Ostadnezhad, S., Delband, M., & Hasanzadeh, A. 2014, NewA, 31, 14

Özdemir, S., Demircan, O., Çiçek, C., & Erdem, A. 2004, AN, 325, 332

Özdemir, S., Demircan, O., Erdem, A. et al. 2002, A&A, 387, 240

Özkardeş, B., & Erdem, A. 2012, NewA, 17, 143

Özkardeş, B., Erdem, A., & Bakış, V. 2009, NewA, 14, 461

Papageorgiou, A., & Christopoulou, P.-E. 2015, AJ, 149, 168

Papageorgiou, A., Christopoulou, P.-E., Pribulla, T., & Vaňko, M. 2015, Ap&SS, 357, 59

Park, J.-H., Lee, J.-W., Kim, S.-L., Lee, C.-U., & Jeon, Y.-B. 2013, PASJ, 65, 1

Pazhouhesh, R., & Edalati, M. T. 2002, IBVS, 5236, 1

Pecaut, M. J., & Mamajek, E. E. 2013, ApJS, 208, 9

Pi, Q.-F., Zhang, L.-Y. Bi, S.-L., Han, X. L., Wang, D.-M., & Lu, H.-P. 2017, AJ, 154, 260

Plewa, T., Haber, G., Wlodarczyk, K. J., & Krzesinski, J. 1991, AcA, 41, 291

Pribulla, T., Chochol, D., Vanko, M., & Parimucha, S. 2002, IBVS, 5258, 1

Pribulla, T., Kreiner, J. M., & Tremko, J. 2003a, CoSka, 33, 38

Pribulla, T., Parimucha, S., Chochol, D., & Vanko, M. 2003b, IBVS, 5414, 1

Pribulla, T., Rucinski, S. M., Blake, R. M. et al. 2009a, AJ, 137, 3655

Pribulla, T., Rucinski, S. M., DeBond, H. et al. 2009b, AJ, 137, 3646

Pribulla, T., Rucinski, S. M., Lu, W.-X. et al. 2006, AJ, 132, 769

Pribulla, T., Rucinski, S. M., Conidis, G. et al. 2007, AJ, 133, 1977

Pribulla, T., & Vanko, M. 2002, CoSka, 32, 79

Pribulla, T., Vanko, M., Chochol, D., Hambálek, L., & Parimucha, Š. 2011, AN, 332, 607

Pribulla, T., Vanko, M., Chochol, D., & Parimucha, S. 2001, CoSka, 31, 26

Pych, W., Rucinski, S. M., DeBond, H. et al. 2004, AJ, 127, 1712

Qian, S.-B. 2001, MNRAS, 328, 914

Qian, S.-B. 2003, MNRAS, 342, 1260

Qian, S.-B., He, J.-J., Liu, L., Zhu, L.-Y., & Liao, W.-P. 2008, AJ, 136, 2493

Qian, S.-B., Liu, L., Soonthornthum, B., Zhu, L.-Y., & He, J.-J. 2007a, AJ, 134, 1475

Qian, S.-B., Wang, J.-J., Zhu, L.-Y., Snoonthornthum, B., Wang, L.-Z., Zhao, E. G., Zhou, X., Liao, W.-P., & Liu, N.-P. 2014, ApJS, 212, 4

Qian, S.-B., Xiang, F.-Y., Zhu, L.-Y., Dai, Z.-B., He, J.-J., & Yuan, J.-Z. 2007b, AJ, 133, 357

Qian, S.-B., & Yang, Y.-G. 2004,AJ, 128, 2430

Qian, S.-B., & Yang, Y.-G. 2005, MNRAS, 356, 765

Qian, S.-B., Yang, Y.-G., Soonthornthum, B., Zhu, L.-Y., He, J.-J., & Yuan, J.-Z. 2005, AJ, 130, 224

Qian, S.-B., Yuan, J.-Z., Xiang, F.-Y., Soonthornthum, B., Zhu, L.-Y., & He, J.-J. 2007c, AJ, 134, 1769

Rafert, J. B., Markworth, N. L., & Michaels, E. J. 1985, PASP, 97, 310

Rainger, P. P., Bell, S. A., & Hilditch, R. W. 1990, MNRAS, 246,47

Rasio, F. A. 1995, ApJ, 444, 41

Ricker, G. R., et al. 2015, J. Astron. Telescopes, Inst. Syst., 1, 014003

Robertson, J. A., & Eggleton, P. P. 1977, MNRAS, 179, 359

Rodríguez, E., Claret, A., Garca, J. M. et al. 1998, A&A, 336, 920

Rucinski, S. M. 1992, AJ, 103, 960

Rucinski, S. M., Baade, D., Lu, W. X., & Udalski, A. 1992, AJ, 103, 573

Rucinski, S. M., Capobianco, C. C., Lu, W.-X. et al. 2003, AJ, 125, 3258

Rucinski, S. M., & Duerbeck, H. W. 2006, AJ, 132, 1539

Rucinski, S. M., & Lu, W.-X. 1999, AJ, 118, 2451

Rucinski, S. M., Lu, W.-X., Capobianco, C. C. et al. 2002, AJ, 124, 1738

Rucinski, S. M., Lu, W.-X., & Mochnacki, S, W. 2000, AJ, 120, 1133

Rucinski, S. M., Lu, W.-X., Mochnacki, S, W., Ogłoza, W., & Stachowski, G. 2001, AJ, 122, 1974

Rucinski, S. M., Pribulla, T., Mochnacki, S. W. et al. 2008, AJ, 136, 586

Rucinski, S. M., Pych, W., Ogłoza, W. et al. 2005, AJ, 130, 767

Russo, G., Sollazzo, C., Maceroni, C., & Milano, L. 1982, A&AS, 47, 211

Samec, R. G., Faulkner, D. R., &Williams, D. B. 2004, AJ, 128, 2997

Samec, R. G., Fuller, R. E., & Kaitchuck, R. H. 1989a, AJ, 97, 1159

Samec, R. G., & Hube, D. P. 1991, AJ, 102, 1171

Samec, R.G., Pauley, B. R., & Carrigan, B. J. 1997, AJ, 113, 401

Samec, R. G., Tuttle, J. P., Brougher, J. A., Moore, J. E., & Faulkner, D. R. 1999, IBVS, 4811,1

Samec, R, G., van Hamme, W., & Bookmyer, B. B. 1989b, AJ, 98, 2287

Samus, N. N., Kazarovets, E. V., Durlevich, O. V., Kireeva, N. N., & Pastukhova, E. N. 2017, ARep, 61, 80

Sarotsakulchai, T., Qian, S.-B., Soonthornthum, B. et al. 2018, AJ, 156, 199

Sarotsakulchai, T., Qian, S.-B., Soonthornthum, B. et al. 2019, PASJ, 71, 34

Schieven, G., Morton, J. C., McLean, B. J., & Hughes, V. A. 1983, A&AS, 52, 463

Schlafly, E. F. & Finkbeiner, D. P. 2011, ApJ 737, 103

Selam, S., Albayrak, B., Yilmaz, M., Şenavci, H. V., Özavci, I., & Ҳetintaş, C. 2005, Ap&SS, 296, 305

Selam, S. O., Esmer, E. M., Şenavcı, H. V. et al. 2018, Ap&SS, 363, 34

Şenavcı, H. V., Nelson, R. H., Özavcı, İ., Selam, S. O., & Albayrak, B. 2008, NewA, 13, 468

Sezer, C., Gulmen, O., & Gudur, N. 1985, Ap&SS, 115, 309

Siwak, M., Zola, S., & Koziel-Wierzbowska, D. 2010, AcA, 6, 305

Shappee, B. J., Prieto, J. L., Grupe, D. et al. 2014, ApJ, 788, 48

Stepien, K. 2006, AcA, 56, 347

Stepien, K. 2011, AcA, 61, 139

Szalai, T., Kiss, L. L., Mészáros, Sz., Vinkó, J., & Csizmadia, Sz. 2007, A&A, 465, 943

Tas, G., & Evren, S. 2006, IBVS, 5687, 1

Tian, X.-M., Zhu, L.-Y., Qian, S.-B., Li, L.-J., & Jiang, L.-Q. 2018, RAA, 18, 20

Terrell, D., & Wilson, R. E., 2005, Ap&SS, 296, 221

Ulaş, B., Kalomeni, B., Keskin, V., Köse, O., & Yakut, K. 2012, NewA, 17, 46

Ulaş, B., & Ulusoy, C. 2014, NewA, 31, 56

Van Hamme, W. 1993, AJ, 106, 2096

Vanko, M., Pribulla, T., Chochol, D., Parimucha, S., Kim, C.-H., Lee, J.-W., & Han, J.-Y. 2001, CoSka, 31,129

Verrot, J. P., & van Cauteren, P. 2000, IBVS, 4910, 1

Vinko, J., Hegedus, T., & Hendry, P. D. 1996, MNRAS, 280,489

Twigg, L. W. 1979, MNRAS, 189, 907

Wadhwa, S. S., & Zealey, W. J. 2005, Ap&SS, 295, 463

Walker, R. L. 1973, IBVS, 855, 1

Wang, J. -J., Jiang, L. -Q., Zhang, B., Zhao, S. -Q., & Yu, J. 2017, PASP, 129, 124202

Wang, K. 2017, RAA, 17, 112

Wang, Y.-R., & Lu, W.-X. 1990, AcApS, 10, 248

Wang, K., Zhang, X.-B., Deng, L.-C. et al. 2015, ApJ, 805, 22

Wilson, R. E., & Devinney, E. J. 1971, ApJ, 166, 605

Wilson, R. E. 1978, ApJ, 224, 885

Wilson, R. E. 1979, ApJ, 234, 1054

Wilson, R. E. 1990, ApJ, 356, 613

Wilson, R. E. 1994, PASP, 106, 921

Wolf, M., Molʼık, P., Hornoch, K., & Šarounová, L. 2000, A&AS, 147, 243

Xiang, F.-Y., Yu, Y.-X., & Xiao, T.-Y. 2015, AJ, 149, 62

Yamasaki, A., Jugaku, J., & Seki, M. 1988, AJ, 95, 894

Yang, Y.-G. 2012, RAA, 12, 419

Yang, Y.-G., Qian, S.-B., & Koppelman, M. D. 2005a, ChJAA, 5, 137

Yang, Y.-G., Qian, S.-B., & Soonthornthum, B. 2012, AJ, 143, 122

Yang, Y.-G., Qian, S.-B., & Zhu, C.-H. 2004, PASP, 116, 826

Yang, Y.-G., Qian, S.-B., & Zhu, L.-Y. 2005b, AJ, 130, 2252

Yang, Y.-G., Qian, S.-B., Zhu, L.-Y., Dai, H.-F., & Soonthornthum, B. 2013, AJ, 146, 35

Yang, Y.-G., Qian, S.-B., Zhu, L.-Y., & He, J.-J. 2009, AJ, 138, 540

Yang, Y.-G., Qian, S.-B., Zhu, L.-Y., He, J.-J., & Yuan, J.-Z. 2005c, PASJ, 57, 983

Yang, Y.-G., Yuan, H.-Y., & Dai, H.-F. 2019AJ....157..111

Yang, Y., & Liu, Q. 1999, A&AS, 136, 139

Yang, Y., & Liu, Q. 2003a, A&A, 401, 631

Yang, Y., & Liu, Q. 2003b, AJ, 126, 1960

Yang, Y., & Liu, Q. 2003c, NewA, 8, 465

Yang, Y., & Liu, Q. 2004, ChJAA, 4, 553

Yakut, K., & Eggleton, P. P. 2005, ApJ, 629, 1055

Yakut, K., Kalomeni, B., & İbanoğlu, C. 2004, A&A, 417, 725

Yakut, K., İbanoğlu, C., Kalomeni, B., & Değirmenci, Ö. L. 2003, A&A, 401, 1095

Yaşarsoy, B., & Yakut, K. 2013, AJ, 145, 9

Yildirim, M. F., Aliçavuş, F., & Soydugan, F. 2019, RAA, 19, 10

Yılmaz, M., Baştürk, Ö., Özavcı, İ., Şenavcı, H. V., & Selam, S. O. 2015, NewA, 34, 271

Yue, Q., Zhang, L.-Y., Han, X. L., Lu, H.-P., Long, L., & Yan, Y. 2019, RAA, 19, 97

Zasche, P., & Uhlář, R. 2010, A&A, 519, 78

Zhai, D.-S., & Lu, W.-X. 1988, AcASn, 29, 9

Zhang, L.-Y., Zhu, Z.-Z., Yue, Q. et al. 2020a, MNRAS, 491, 6065

Zhang, X.-D., Qian, S.-B., & Liao, W.-P. 2020b, MNRAS, 492, 4112

Zhang, J., Qian, S.-B., Han, Z.-T., & Wu Y. 2017, MNRAS, 466, 1118

Zhang, J.-T., Zhang, R.-X., & Zhai, D.-S. 1992, AcASn, 33, 131

Zhang, X. B., Deng, L., & Lu, P. 2009, AJ, 138, 680

Zhang, X. B., & Zhang, R. X. 2004, MNRAS, 347,307

Zhou, X., Qian, S.-B., He, J.-J., Zhang, J., Jiag, L.-Q. 2015, PASJ, 67, 98

Zhou, X., Qian, S.-B., Essam, A., He, J.-J., & Zhang, B. 2016a, NewA, 47, 37

Zhou, X., Qian, S.-B., Zhang, J., Zhang, B., & Kreiner, J. 2016b, AJ, 151, 67

Zhu, L.-Y., Qian, S.-B., Liu, N.-P., Liu, L., Jiang, L.-Q. 2013, AJ, 145, 39

Zola, S., Baran, A., Debski, B., & Jableka, D. 2017, MNRAS, 466, 2488

Zola, S., Kreiner, J. M., Zakrzewski, B. et al. 2005, AcA, 55, 389

Zola, S., Niarchos, P., Manimanis, V., & Dapergolas, A. 2001, A&A, 374, 164

Zola, S., Rucinski, S. M., Baran, A. et al. 2004, AcA, 54, 299

Zwitter, T., Munari, U., Marrese, P. M. et al. 2003, A&A, 404, 333